\documentclass[a4paper, 11pt]{article}
\usepackage{jheppub}

\pdfoutput=1
\usepackage{amsmath,amssymb}
\usepackage{mathtools}

\title{ JIMWLK Evolution, Lindblad Equation and  Quantum-Classical Correspondence}
\author{Ming Li }
\author{and Alex Kovner}
\affiliation{Physics Department, University of Connecticut, Storrs, CT 06269, USA}

\emailAdd{ming.2.li@uconn.edu}
\emailAdd{alexander.kovner@uconn.edu}

\abstract{
In the Color Glass Condensate(CGC) effective theory, the physics of valence gluons with large longitudinal momentum is reflected in  the distribution of color charges in the transverse plane. Averaging over the valence degrees of freedom is effected by integrating over classical color charges with some quasi probability weight functional $W[{\mathbf{j}}]$ whose evolution with rapidity is governed by the JIMWLK  equation.  In this paper, we reformulate this setup in terms of effective quantum field theory on valence Hilbert space governed by the reduced density matrix $\hat\rho$  for hard gluons, which is obtained after properly integrating out the soft gluon ``environment". We  show that the evolution of this density matrix with rapidity in the dense and dilute limits has the form of Lindblad equation. The quasi probability distribution (weight) functional $W$  is  directly related to the reduced density matrix $\hat\rho$ through the generalization of the Wigner-Weyl quantum-classical correspondence, which reformulates quantum dynamics on Hilbert space in terms of classical dynamics on the phase space. In the present case the phase space is non Abelian and is spanned by the components of transverse color charge density ${\mathbf{j}}$. The same correspondence maps the Lindblad equation for $\hat\rho$ into the JIMWLK evolution equation for $W$ .
}

\begin{document}
\maketitle
\flushbottom

\section{Introduction}
This paper examines  the status of the JIMWLK evolution equation \cite{Balitsky:1995ub, JalilianMarian:1997jx, JalilianMarian:1997gr, JalilianMarian:1997dw, Kovner:2000pt, Iancu:2000hn, Ferreiro:2001qy, Mueller:2001uk}  in relation to  the effective density matrix of a high energy hadronic system. We are motivated to address this question by the discussion in a recent paper \cite{Armesto:2019mna} which suggested an extension of JIMWLK framework to include a wider set of observables other than just color charge density $j^a(x)$ in the hadronic wave function. The starting point of \cite{Armesto:2019mna} is the interpretation of JIMWLK evolution equation as the equation for diagonal matrix elements of the density matrix in the color charge density basis. Although this interpretation is natural  when the color charge density is large, it is not quite clear how to formalize it for low density, since in this regime the commutator of the color charge density operators is not negligible and a basis in which all components of $j^a(x)$ are diagonal obviously does not exist.
On the other hand, as shown a while ago \cite{Kovner:2005uw}  the calculations of averages in this regime as well can be formulated in terms of the functional integral over classical fields $j^a(x)$, which suggests that perhaps such interpretation albeit possibly modified, can be put forward after all.

An interesting suggestion of \cite{Armesto:2019mna} is that the rapidity evolution of the generalized CGC density matrix is of the Lindblad type \cite{Gorini:1975nb},\cite{Lindblad:1975ef}. This type of evolution is very general in quantum mechanical systems where one follows only part of the degrees of freedom by reducing the density matrix over the ``environment" (the unobserved degrees of freedom in the Hilbert space). If the 
 ``environment" degrees of freedom have  only short range correlation in time, the dynamics of the observed part of the system is Markovian and is therefore governed by a differential equation. The Lindblad form of such evolution is the most general one that preserves the properties of the density matrix stemming from its probabilistic nature (normalization and positivity of all eigenvalues). 
Although Lindblad equation naturally arises in {\it time evolution} of quantum systems, JIMWLK evolution is of a somewhat different nature. It describes the change of the system with rapidity (or energy) but not in time. It is thus not obvious whether one should expect Lindblad form  to be generic in this context and if yes, under what conditions.

In this paper we try to address these questions. We arrive at two basic results. First, we show that JIMWLK evolution can indeed be understood as evolution of a density matrix. Within the JIMWLK framework however, the density matrix is not generic, but is rather assumed to depend only on the operators $\hat j^a(x)$ which satisfy the standard $SU(N)$ commutation relations. The fact that $\hat\rho$ depends only on the generators of the $SU(N)$ group means that it has a quasi diagonal form - i.e. it does not connect states belonging to different representations of $SU(N)$.   It is in this sense that the reduced density matrix is (almost) diagonal even if the commutators of $j^a$ cannot be neglected. The consequence of this strong assumption on the nature of the density matrix  is that the states in the  Hilbert space of the reduced system are completely specified by their $SU(N)$ transformation properties at every transverse position $x$, and therefore in the reduced space one loses track of the gluon longitudinal momentum as well as polarization. 

Second, we show that on this Hilbert space the JIMWLK evolution is indeed equivalent to Lindblad type equation for this restricted set of density matrices. The same applies to the so called KLWMIJ evolution which describes the dilute regime. In analogy with  time evolution of quantum mechanical systems, the Lindblad equation arises in the situation when the correlations in the ``unobserved" part of the system are short range in rapidity. However we also argue that in general (i.e. away from the dense - JIMWLK and dilute - KLWMIJ limits) the high energy evolution is unlikely to be of Lindblad type. This follows from certain general properties of our derivation of the evolution of the density matrix based on the calculation of the CGC wave function presented in \cite{Altinoluk:2009je} . Although the calculations of \cite{Altinoluk:2009je} are strictly valid only in the aforementioned limits, the general features of the derivation are expected to be more universal. The reason that the Lindblad form is not expected to arise, is that in the high energy evolution framework, the rapidity does not just play the role of the evolution parameter, but also that of the label on the quantum states of the gluons which are integrated out. In this situation in general one does not expect the Lindblad form for the differential equation. Thus to ensure Lindblad form nontrivial conditions on dependence of the matrix elements on gluon rapidities have to be satisfied. We discuss this point in detail in Section IV.

Another result of this paper is the precise mathematical relation between the effective density matrix, and the ``probability density function" $W[{\mathbf j}]$ that usually appears in the literature as the subject of JIMWLK evolution. 
 We confirm that the quantum mechanical averaging with the density matrix $\hat\rho$ is mapped into 
 the calculation of observables in terms of functional integral over {\it classical} fields $j^a(x)$ with the weight functional $W[{\mathbf{j}}]$, as indeed always done in the CGC literature. This functional integral must be regarded as an integral over the {\it phase space} variables of the classical system, and not its {\it configuration space} variables. This quantum-classical correspondence between the quantum density matrix and the classical functional of phase space variables $W[{\mathbf{j}}]$ is deeply related to the correspondence between the density matrix and Wigner function in ordinary quantum mechanics. In the context of high energy evolution we require  a generalization of the original Wigner-Weyl correspondence\cite{Hillery:1983ms} since the phase space of the theory is spanned not by operators $q$ and $p$ which constitute the Heisenberg algebra, but rather by operators $j^a$ with the $SU(N)$ algebra.
Nevertheless the basic correspondence  involves mappings between quantum operators and Hilbert space on one side and classical quantities and phase space on the other side in the sense of Weyl's correspondence rule. The weight functional $W[{\mathbf{j}}]$ is consequently identified as the Wigner functional \cite{Hillery:1983ms} and can indeed be considered as a quasi-probability distribution on the phase space. 

The outline of this paper is the following.  In Sec. II, we give a brief review of the Lindblad equation for density matrix of an open quantum system and a recap of the Hamiltonian formalism of CGC effective theory.  In Sec. III we explain how to define the reduced CGC density matrix, and show that its rapidity evolution has the  Kraus form, which is a general evolution that preserves probabilistic interpretation of a density matrix, but is not  necessarily differential. In Sec. IV we derive the differential evolution of the density matrix using the analog of Markovian porperty, i.e. the fact that the correlation of the ``environment" is short range in rapidity. We show that in the dilute (KLWMIJ) and dense (JIMWLK) limits the differential  evolution equation is indeed of the Lindblad type.  We also discuss the properties of the derivation which suggest that the standard Lindblad form is bound to be modified away from these limits.
To be clear, in this paper we do not go beyond the original JIMWLK setup in the sense that we consider  density matrices that only depend on color charge density operators, and thus presently our derivation does not extend to ideas put forward in \cite{Armesto:2019mna}. In Sec. V,  we derive the explicit relation between the standard JIMWLK approach and the density matrix approach described in this paper. We show that the two are related by a variant of the Wigner-Weyl quantum-classical correspondence and spell out explicitly the correspondence rules which transform one setup into the other. The JIMWLK and KLWMIJ equations are  then reproduced by mapping the Lindblad equation for the density matrix in the appropriate (dense and dilute) limits into the classical phase space. Finally Sec. VI contains a short discussion.

\section{Review of Basics}
\subsection{Lindblad equation for open quantum systems}
In this section we present a short review of Lindblad equation for open systems.

Lindblad equation is the most general Markovian and non-unitary evolution equation for density matrix of an open quantum system. This equation preserves the properties of density matrix: hermiticity, unit trace and positivity.  Here we follow the heuristic discussions by Preskill \cite{Preskill:2019}. More physical derivations and applications can be found in the books \cite{Carmichael:1993,Breuer:2007}. 

Consider a bipartite system involving two subsystems: the ``observed system" and the ``environment" with the Hamiltonians $\hat{H}_s$ and $\hat{H}_e$, respectively. The two subsystems interact via the Hamiltonian  $\hat{H}_{se}$.  The total density matrix of the complete system evolves according to the quantum Liouville equation
\begin{equation}
\frac{d \hat{\rho}}{dt} = -i [\hat{H}, \hat{\rho}]
\end{equation}
with $\hat{H} = \hat{H}_s + \hat{H}_e +\hat{H}_{se}$.  Formally, the solution can be expressed as 
\begin{equation}
\hat{\rho}(t) = \hat{U}(t) \hat{\rho}(0) \hat{U}^{\dagger}(t)
\end{equation}
with $\hat{U}(t) = e^{-i\hat{H} t}$. To obtain the density matrix of the observed subsystem after a finite time, one traces over the Hilbert space of the environment.
Let us assume that the initial total density matrix  is a direct product of the density matrices of the observed system and the environment $\hat{\rho}(0) = \hat{\rho}_s(0) \otimes \hat{\rho}_e(0) = \hat{\rho}_s(0) \otimes |0_e\rangle\langle 0_e|$. For simplicity let us take the environment to be initially in a pure state denoted by $|0_e\rangle$, which can be thought of as the ground state without loss of generality.  The  density matrix of the observed system is then expressed as
\begin{equation}\label{eq:kraus_rep}
\hat{\rho}_s(t) = \mathrm{Tr}_e \hat{\rho} (t) =  \sum_{n}\langle n |\hat{U}(t) | 0_e\rangle \hat{\rho}_s (0) \langle 0_e| \hat{U}^{\dagger}(t) |n\rangle =\sum_n \hat{M}_n(t) \hat{\rho}_s(0) \hat{M}^{\dagger}_n(t).
\end{equation}
Here $\left\{ |n\rangle \right\}$ represents a complete basis in the Hilbert space of the environment. The objects $\hat{M}_n(t) = \langle n |\hat{U}(t) | 0_e\rangle$, sometimes called superoperators,  are operators on the Hilbert space of the observed system and govern the evolution of its density matrix. As far as the dynamics of the environment is considered, $\hat{M}_n(t)$ represents the transition amplitude for the environment, which is initially in the state $|0_e\rangle$, to be in the state $|n\rangle$ after a finite time $t$. They satisfy the property $\sum_n \hat{M}_n^{\dagger}(t)\hat{M}_n(t) =1$ following from the unitarity of $\hat{U}(t)$. 

The time evolution of density matrix in Eq.\eqref{eq:kraus_rep} has been expressed in an operator summation form which is also called a Kraus representation. It is easy to check that the Kraus representation  preserves the hermiticity, unit trace and positivity of the density matrix.  It is believed that any reasonable time evolution of density matrices should have a Kraus representation. 

The general  Kraus representation Eq.\eqref{eq:kraus_rep} does not have the form of a differential equation for the evolution of the  density matrix. It is only under the Markovian approximation that an equivalent expression in terms of a differential equation becomes possible. 
The Markovian approximation holds if the typical correlation time between the environment degrees of freedom $t_{corr}$ is shorter than the typical inverse frequency of the observed system $\Delta t_s$, which is of the order of the relevant ``discretization" time step for approximate differential time evolution. If this is the case the state of environment is only affected by the state of the observed system at the particular time of observation (measured with accuracy $\Delta t_s$), and thus the back reaction - the effect of the environment on the observed system is local in time. We note that this is the typical Born-Oppenheimer situation, when the environment is associated with fast degrees of freedom, while the observed system is relatively slow. In the opposite regime it is clear that local (differential) time evolution is impossible, since the backreaction of the environment on the system will depend on the state of the system at some past time.

In Markovian regime one then proceeds as follows. For an infinitesimal period of time, only terms linear in $dt$ should be kept on the right hand side of Eq. \eqref{eq:kraus_rep}. 
The superoperators for $n>0$, have the structure
\begin{equation}\label{amp}
\hat{M}_n(dt) =  \sqrt{dt}\hat{L}_n, \quad n>0
\end{equation}
The argument here is that $\hat{M}^\dagger_n(t)\hat{M}_n(t)$ is the probability for the environment to "jump" to the state $n$ during the time $t$. For small enough $t$ (but such that $t>t_{corr}$) this probability should grow linearly with $t$. 
The operators $\hat L_n$ are called Lindblad operators or jump operators as they involve transitions of the environment to different states after an infinitesimal time.

The remaining superoperator has the form
\begin{equation}
\hat{M}_0(dt) = 1 + (-i \hat{H}_s + \hat{K}) dt
\end{equation}
with $H_s$ and $K$ being Hermitian. This is the transition amplitude for the environment to be in its original state after an infinitesimal time and should be linear in time for small enough times. The operator $\hat{K}$ is related to the wave function renormalization effect and $\hat{H}_s$ governs the unitary evolution of the system without causing any changes to the environment.


The Kraus normalization condition $\sum_n\hat{M}^{\dagger}_n(dt) \hat{M}_n (dt)=1$ relates the wavefunction renormalization operator $\hat{K}$ to the jump operators by
\begin{equation}
\hat{K} = -\frac{1}{2}\sum_{n>0} \hat{L}^{\dagger}_n \hat{L}_n\, .
\end{equation}
Taking the limit $dt\rightarrow 0$, the Kraus representation then becomes an differential equation
\begin{equation}
\frac{d\hat{\rho}_s}{dt} = -i[\hat{H}_s, \hat{\rho}_s]  + \sum_{n>0} \left( \hat{L}_n\hat{\rho}_s \hat{L}^{\dagger}_n -\frac{1}{2}\hat{L}_n^{\dagger}\hat{L}_n \hat{\rho}_s-\frac{1}{2} \hat{\rho}_s \hat{L}_n^{\dagger}\hat{L}_n \right)\, .
\end{equation}
This is the Lindblad equation, or sometimes known as Gorini-Kossakowski-Lindblad-Sudarshan master equation \cite{Gorini:1975nb,Lindblad:1975ef}.

 \subsection{The soft gluon vacuum and the CGC}
 We now review the derivation of the high energy evolution\cite{Kovchegov:2012mbw} .
 There exist two equivalent approaches to the derivation  of the CGC effective theory. One is based on the Lagrangian formalism \cite{McLerran:1993ni, McLerran:1993ka, JalilianMarian:1997jx, JalilianMarian:1997gr, JalilianMarian:1997dw, Iancu:2000hn, Ferreiro:2001qy} and the other on the Hamiltonian formalism \cite{Kovner:2005pe, Kovner:2007zu}. We briefly review the Hamiltonian formalism as it will be the starting point for deriving the Lindblad equation for the CGC density matrix.
 
 The derivation of the JIMWLK evolution equation starts with establishing the ground state wave function of soft gluon modes in the background of more energetic gluons which are described by a  color charge density field.
 
  In the light cone gauge $A^+=0$, the Hamiltonian of the pure gluonic sector of QCD is
\begin{equation}
H = \int dx^-d\mathbf{x}_{\perp} \left(\frac{1}{2} \Pi^-_a(x^-, \mathbf{x}_{\perp}) \Pi^-_a(x^-,\mathbf{x}_{\perp}) + \frac{1}{4}F_{ij}^a(x^-,\mathbf{x}_{\perp}) F^a_{ij}(x^-,\mathbf{x}_{\perp})\right) 
\end{equation}
with the chromoelectric and chromomagnetic parts being
\begin{equation}
\begin{split}
&\Pi^-_a(x^-, \mathbf{x}_{\perp}) = \partial_-A^-_a(x^-,\mathbf{x}_{\perp})=\frac{1}{\partial_-}\left( D_i^{ab} \partial_-A_i^b(x^-,\mathbf{x}_{\perp})\right)\, ,\\
&F_{ij}^a(x^-,\mathbf{x}_{\perp})  = \partial_iA_j^a(x^-,\mathbf{x}_{\perp}) -\partial_j A_i(x^-,\mathbf{x}_{\perp}) -gf^{abc}A_i^b(x^-,\mathbf{x}_{\perp})A_j^b(x^-,\mathbf{x}_{\perp})
\end{split}
\end{equation}
 The covariant derivative is defined as $D^{ab}_i = \partial_i\delta^{ab} -gf^{acb} A_i^c$ and $\partial_- = \partial/\partial x^-$ is the longitudinal spatial derivative.  The $1/\partial_-$ operator in the expression of the chromoelectric field has to be regularized as it contains a singularity at vanishing longitudinal momentum, $k^+=0$. This singularity is ultimately related to the zero mode in the $A_i^a(x^-,\mathbf{x}_{\perp})$ fields and is regulated by imposing a residual gauge fixing condition. We choose the residual gauge fixing 
 \begin{equation}\label{eq:gauge_fixing}
 \partial_i A_i^a(x^-\rightarrow-\infty) =0\, 
 \end{equation}
 
 One separates the gluonic degrees of freedom imposing a longitudinal momentum separation scale $\Lambda^+$. In the high energy limit, the dominant interaction between soft gluons ($k^+<\Lambda^+$) and hard gluons ($k^+>\Lambda^+$) has the form of eikonal coupling $A^-_a J_a^+$ with $J^+_a$ representing the color charge density of the hard gluons and $A^-_a$ representing soft gluons.  This interaction term emerges from the chromoelectric part of the Hamiltonian and involves the specific expressions $J^+_a = -gf^{abc}A_i^b\partial_-A_i^c$ and $A^-_a = \frac{1}{\partial_-}\Pi^-_a$. Furthermore, as far as soft gluons are concerned, the hard gluon dynamics can be viewed as frozen in time so that the color current $J^+_a \equiv J^+_a(x^-,\mathbf{x}_{\perp})$ is time independent at the lowest order. All in all, the Hamiltonian for the soft gluonic modes  becomes
 \begin{equation}\label{eq:cgc_hamiltonian}
H_{CGC} = \int dx^-d\mathbf{x}_{\perp} \left(\frac{1}{2} (\Pi^-_a(x^-, \mathbf{x}_{\perp}) + \frac{1}{\partial_-} J^+_a)^2 + \frac{1}{4}F_{ij}^a(x^-,\mathbf{x}_{\perp}) F^a_{ij}(x^-,\mathbf{x}_{\perp})\right) 
\end{equation}

 Canonical quantization 
 is implemented by promoting the \textit{normal modes} of the full $A_i^a$ fields to operators and imposing the equal (light cone) time commutation relation 
 \begin{equation}\label{eq:commutation_fields}
 [\hat{A}_i^a(x^-, \mathbf{x}_{\perp}), \hat{A}_j^b(y^-,\mathbf{y}_{\perp})] = -\frac{i}{2} \epsilon(x^- - y^-) \delta^{ab}\delta_{ij}\delta(\mathbf{x}_{\perp} - \mathbf{y}_{\perp})
 \end{equation} 
 with the sign function defined as $\epsilon(x) = \frac{1}{2}(\Theta(x) - \Theta(-x))$. In terms of the canonical creation and annihilation operators, the normal modes $\hat{A}_i^a$ have the expansion
 \begin{equation}
 \hat{A}_i^a(x^-, \mathbf{x}_{\perp}) = \int_0^{+\infty} \frac{dk^+}{2\pi} \frac{1}{\sqrt{2k^+}} \left( \hat{a}_i^a(k^+, \mathbf{x}_{\perp}) e^{-ik^+x^-} + \hat{a}^{\dagger a }_i(k^+, \mathbf{x}_{\perp}) e^{ik^+x^-}\right)
 \end{equation}
 with
 \begin{equation}
 \left[ \hat{a}_i^a(k^+,\mathbf{x}_{\perp}), \hat{a}_j^{\dagger b}(p^+, \mathbf{y}_{\perp})\right] = (2\pi) \delta^{ab}\delta_{ij}\delta(k^+-p^+)\delta(\mathbf{x}_{\perp} -\mathbf{y}_{\perp})\, .
 \end{equation}
 The color charges in the leading order are taken to have the extreme Lorentz contracted form $J^+_a(x^-,\mathbf{x}_{\perp}) = \delta(x^-)j^a(\mathbf{x}_{\perp})$ with the transverse color charge density
 \begin{equation}\label{eq:color_current}
\hat{ j}^a(\mathbf{x}_{\perp}) = igf^{abc}\int_{k^+>\Lambda^+} \frac{dk^+}{2\pi} \hat{a}_i^{\dagger b}(k^+,\mathbf{x}_{\perp}) \hat{a}_i^c(k^+,\mathbf{x}_{\perp})\, 
 \end{equation}
 The components of color charge satisfy the commutation relations of the $SU(N)$ algebra 
 \begin{equation}\label{eq:commutation_current}
 \left[\hat{j}^a(\mathbf{x}_{\perp}),\hat{j}^b(\mathbf{y}_{\perp}) \right] = igf^{abc} \hat{j}^c(\mathbf{x}_{\perp})\delta(\mathbf{x}_{\perp} -\mathbf{y}_{\perp})\, .
 \end{equation}
 The Hamiltonian system Eqs. \eqref{eq:cgc_hamiltonian},  \eqref{eq:gauge_fixing}, 
 together with the commutation relations Eqs. \eqref{eq:commutation_fields}, \eqref{eq:commutation_current} constitute the starting point for the derivation of the CGC effective theory. 
 
 The first goal is to find the ground state wave function of the soft glue.  This is in general a very complicated problem, but is simplifies in two parametric regimes. One interesting regime is when the   color charge density is small $j^a\sim g$ (dilute limit).  Here one can treat the interaction with color charges perturbatively. The other regime is  the dense limit  where the color charge is parametrically large $j^a \sim \frac{1}{g}$. Here the simplification is that the commutator of the color charges is (almost) negligible and they can be treated as (almost) classical fields.

  In the dilute limit, the ground state wave function can be found by a direct perturbative calculation.
  The resulting vacuum wave function can be written as
   \begin{equation}
 |\psi_0\rangle  = \hat{\mathcal{C}} |0\rangle 
 \end{equation}
 with the coherent operator 
 \begin{equation}\label{eq:coherent_operator}
 \hat{\mathcal{C}} = \mathrm{Exp}\left\{i\int d\mathbf{x}_{\perp} b_i^a(\mathbf{x}_{\perp}) \int^{\Lambda^+}_{\Lambda^+e^{-\Delta y}} \frac{dk^+}{\pi |k^+|^{1/2}}\left(\hat{a}_i^{\dagger a}(k^+, \mathbf{x}_{\perp}) + \hat{a}_i^{a}(k^+, \mathbf{x}_{\perp})\right)\right\}
 \end{equation}
 where $E$ is the energy of the process.
 
 Here $b_i^a(\mathbf{x}_{\perp})$ satisfy the equations
 \begin{equation}
 \begin{split}
& \partial_i b_i^a(\mathbf{x}_{\perp}) = j^a(\mathbf{x}_{\perp})\, ,\\
&\partial_i b_j^a(\mathbf{x}_{\perp}) - \partial_j b_i^a(\mathbf{x}_{\perp}) -gf^{abc} b_i^b(\mathbf{x}_{\perp}) b_j^c(\mathbf{x}_{\perp}) =0 \, .\\
\end{split}
 \end{equation}
 or, in the dilute regime
 \begin{equation}
 b_i^a(\mathbf{x}_{\perp})=\int d^2y \frac{\partial_i}{\partial^2}(x,y)j^a(y)
 \end{equation}
 In the dense limit, similar analysis applies except now $b_i^a(\mathbf{x}_{\perp})\sim 1/g$ and additional order $\mathcal{O}(1)$ quantum fluctuations on top of the $b_i^a$ fields need to be considered.  One can still use perturbative expansion in $g$, but resumming terms of order $gb$. In the leading order  the Hamiltonian is diagonalized by a nontrivial  Bogoliubov transformation. The detailed analysis appears in \cite{Kovner:2007zu}.  The resulting ground state wavefunction is
 \begin{equation}
 |\psi_0\rangle  = \hat{\mathcal{C}} \hat{\mathcal{B}} |0\rangle \, .
 \end{equation}
 The additional Bogoliubov operator $\hat{\mathcal{B}}$ can be formally expressed as
 \begin{equation}\label{eq:bogoliubov_operator}
 \hat{\mathcal{B}} = \mathrm{Exp}\left\{ \hat{a}_{\alpha}^{\dagger} \Lambda_{\alpha\beta} \hat{a}_{\beta}^{\dagger} +\hat{a}_{\alpha}\Lambda_{\alpha\beta}^{\ast} \hat{a}_{\beta} \right\}\, .
 \end{equation}
 Here $\alpha, \beta$ represent all the possible indices (color, spatial coordinates, polarization, and longitudinal momentum which varies between $\Lambda^+e^{-\Delta y}$ and $\Lambda^+$). The explicit expression of the symmetric matrix $M_{\alpha\beta}$ is not available, however, the action of the Bogoliubov operator on the fundamental degrees of freedom $\hat{A}_i^a$ and $\hat{j}^a$ have been derived. 
 
 The nontrivial structure of the soft gluon ground state leads to appearance of induced color charge density due to the soft gluons modes. This additional color charge density serves as an additional source for even softer gluons which arise in the evolution to even higher rapidities. This is the basic physics of the high energy evolution.

 \section{The Reduced CGC Density Matrix and Its Evolution}
 Having found the vacuum of the soft gluons, we can now address the evolution at high energy.
 We take here a different perspective on this derivation than given in the literature, and discuss the evolution from the point of view of quantum density matrix.

 Given that we have separated our degrees of freedom into soft and hard gluons, we can view our system naturally as bipartite. 
 At some initial rapidity, the soft gluons are in the perturbative vacuum state, and thus  the total density matrix is separable 
 \begin{equation}
 \hat{\rho} = \hat\rho_v \otimes |0\rangle \langle 0|
 \end{equation}
 where  the density matrix $\hat\rho_v$ is an operator on the hard gluon Hilbert space. 
 
 The assumption inherent in the derivation of the JIMWLK equation is that the only relevant degrees of freedom on this Hilbert space are components of the color charge density $\hat j^a({\mathbf x}_\perp)$. This is a crucial assumption. If the valence Hilbert space could be factorized into a direct product of the space spanned by $\hat j^a({\mathbf x}_\perp)$ and  its complement, reducing  over the complement would rigorously define $\hat\rho_v[{\mathbf j}]$. However the full Hilbert space of the valence modes does not have such a direct product structure. It is thus not clear whether a well defined mathematical procedure of ``integrating out"  exists which may reduce the  density matrix so that in general it depends only on $\hat j^a({\mathbf x}_\perp)$. Nevertheless one can simply assume that at initial rapidity the density matrix indeed has such a form. It is then  true (as we will see below) that this form persists throughout the evolution to higher rapidities. We will thus abide by this assumption and will treat $\hat\rho_v$ as an operator that depends only on $\hat j^a({\mathbf x}_\perp)$.

 After boosting the system by a finite rapidity $\Delta y$, the total density matrix changes due to the emission of soft gluons into the newly opened rapidity interval.
 \begin{equation}
 \hat{\rho}(\Delta y) =   \hat{\Omega} |0\rangle \hat{\rho}_v \langle 0|\hat{\Omega}^{\dagger}\, .
 \end{equation}
 The gluon emission operator as discussed above can be written as
 \begin{equation}
 \hat{\Omega} \equiv \Omega [\hat{j}^a, \hat{a}_i^{a \dagger}, \hat{a}_i^a; \Delta y] = \hat{\mathcal{C}}\hat{\mathcal{B}}
 \end{equation} 
 with $\hat{\mathcal{C}}$ and $\hat{\mathcal{B}}$ defined in eqs. \eqref{eq:coherent_operator} and \eqref{eq:bogoliubov_operator}, respectively. This form applies both for the dilute and dense regime of the evolution. Note that $\hat{\Omega} $ depends on the soft gluon creation and annihilation operators as well as the color charge density operator. While the $\hat{a}_i^{a \dagger}, \hat{a}_i^a$ act on the soft vacuum state $|0\rangle$, the $\hat{j}^a$  acts  on the valence (hard) density matrix $\hat\rho_v$. Dependence on $\Delta y$ of $\hat{\Omega}$ is crucial in obtaining the evolution equation. This point will be elaborated in the following.  
 
 Our next goal is to derive the reduced density matrix by tracing over the ``environment" degrees of freedom.   The purpose of this reduction of the Hilbert space is to integrate out all the additional degrees of freedom associated with soft gluons that emerged after boosting the wave function, {\it except} the additional color charge density that they contribute. The reason for this exception is, that  in the next step in the evolution the even softer gluons will couple to the total color charge density, including that due to gluons in the rapidity interval between $y$ and $y+\Delta y$. Our current soft gluons give a nontrivial contribution to this charge density, and we have to keep this extra contribution explicitly, rather than integrate it out.  
 \subsection{Defining the charge shift operator}
 
 Put in different words, we are interested in a general set of observables that depend on rapidity integrated color charge density. Before evolution those are averages of the form
 \begin{equation}
 \langle O(\hat j^a)\rangle={\rm Tr}[O(\hat j^a)\hat\rho_v]
 \end{equation}
 while after a step $\Delta y$ of the evolution
 \begin{equation}\label{oav}
  \langle O(\hat j^a+\hat j^a_{\mathrm{soft}})\rangle={\rm Tr}[O(\hat j^a+\hat j^a_{\mathrm{soft}})\hat\rho(\Delta y)]
 \end{equation}
 Here $\hat{j}^a_{\mathrm{soft}}(\mathbf{x}_{\perp})$ has the explicit expression eq.\eqref{eq:color_current} with the longitudinal momentum integration restricted in the rapidity range $\Delta y$. 
 
 It is thus clear that we should not simply reduce the density matrix over the Hilbert space of soft gluons,  but ``partially" trace over the soft gluons integrating out all degrees of freedom except the color charge density.
 To facilitate this partial tracing over soft gluons, we introduce the operator $\hat{R}$,  which is defined by its action on $\hat{j}^a$, 
 \begin{equation}\label{rj}
 \hat{R}^{\dagger} \hat{j}^a(\mathbf{x}_{\perp})\hat{R} = \hat{j}^a(\mathbf{x}_{\perp}) + \hat{j}^a_{\mathrm{soft}}(\mathbf{x}_{\perp})\, .
 \end{equation}
 so that 
  \begin{equation}\label{oj}
 \hat{R}^{\dagger}O( \hat{j}^a(\mathbf{x}_{\perp}))\hat{R} = O(\hat{j}^a(\mathbf{x}_{\perp}) + \hat{j}^a_{\mathrm{soft}}(\mathbf{x}_{\perp}))\, .
 \end{equation}
 for any operator $O$.
  It may not be obvious that  $\hat R$ can be properly defined as an operator on the Hilbert space, given that different components of $\hat{j}^a(x)$ are noncommuting operators. As we show now, this nevertheless is the case.
  
  Let us introduce the operator $\hat{\Phi}^a(\mathbf{x}_\perp)$ via
   \begin{equation}\label{R}\hat{R} = \mathrm{Exp}\{ - i\int d^2\mathrm{x}_{\perp} \hat{j}^a_{\mathrm{soft}}(\mathbf{x}_{\perp})\hat{\Phi}^a(\mathbf{x}_{\perp})\}\end{equation}

 We will look for  $\hat{\Phi}^a$ (we omit the transverse coordinate dependence for simplicity) as a set of operators acting on the same Hilbert space as $\hat j^a$ satisfying the following commutation relations
\begin{equation}\label{com}
\begin{split}
&[\hat{\Phi}^a, \hat{\Phi}^b] =0\, ,\\
&[\hat{\Phi}^a, \hat{j}^b] = M^{ab}(\hat\Phi)\, \\
\end{split}
\end{equation}
with $M$ chosen to satisfy the requirement 
\begin{equation}\label{eq:shift_requirement}
\mathrm{exp}\left\{i\hat{j}^a_{\mathrm{soft}} \hat{\Phi}^a\right\} \hat{j}^e\, \mathrm{exp}\left\{-i\hat{j}^a_{\mathrm{soft}} \hat{\Phi}^a\right\} = \hat{j}^e + \hat{j}^e_{\mathrm{soft}}\, .
\end{equation}
In calculating the action of $\hat R$ we assume that the operators $\hat j^a$ satisfy $SU(N)$ algebra, and so do the operators  $\hat j^a_{\mathrm{soft}}$, while the two set of operators commute with each other.

We use the Baker-Hausdorff formula
\begin{equation}
e^X Y e^{-X} = Y + [X, Y] + \frac{1}{2!}[X, [X, Y]] + \frac{1}{3!}[X,[X,[X,Y]]]+\ldots + \frac{1}{n!}[X, [X, [\ldots [X, Y]\ldots]]] + \ldots
\end{equation}
With the commutation relations eq.(\ref{com}) we have (for adjoint representation $-if^{abc} = T^a_{bc}$)
\begin{equation}
[i\hat{j}^a_{\mathrm{soft}} \hat{\Phi}^a, \hat{j}^e]   = i\hat{j}^a_{\mathrm{soft}} M^{ae}(\hat\Phi)
\end{equation}
\begin{equation}
\frac{1}{2!}[i\hat{j}^b_{\mathrm{soft}} \hat{\Phi}^b, [i\hat{j}^a_{\mathrm{soft}} \hat{\Phi}^a, \hat{j}^e]   ]= \frac{1}{2!} \, i\hat{j}^a_{\mathrm{soft}} \left(igT^b \hat{\Phi}^b M(\hat\Phi)\right)_{ae}
%
\end{equation}
\begin{equation}
\frac{1}{3!}[i\hat{j}^c_{\mathrm{soft}} \hat{\Phi}^c,[i\hat{j}^b_{\mathrm{soft}} \hat{\Phi}^b, [i\hat{j}^a_{\mathrm{soft}} \hat{\Phi}^a, \hat{j}^e]   ] ] = \frac{1}{3!} i\hat{j}^a_{\mathrm{soft}} \left((igT^b \hat{\Phi}^b)^2 M(\hat\Phi)\right)_{ae}
\end{equation}
Let us  take the ansatz
\begin{equation}
M^{ab}(\hat\Phi) = -i\sum_{n=0}^{\infty} c_n \left[\chi^n\right]_{ab}\, , \quad\mathrm{with}\,\,\,  \chi = igT^b\hat{\Phi}^b\, .
\end{equation}
Clearly, $c_0 =1$ follows from the requirement eq. \eqref{eq:shift_requirement}. This requirement further imposes the constraint
\begin{equation}
i + \sum_{k=0}^{\infty} \frac{1}{(k+1)!} \left[\chi^k M(\chi)\right]^{ab}=0
\end{equation}
which after substituting the ansatz for $M(\chi)$ becomes
\begin{equation}
1= \sum_{k=0}^{\infty}\sum_{m=0}^{\infty}\frac{c_m}{(k+1)!} \chi^{k+m}_{ab}
\end{equation}
which is equivalent to the following recursive relations 
\begin{equation}\label{eq:recursive_relations}
c_N = -\sum_{m=0}^{N-1} \frac{c_m}{(N-m+1)!} \, , \quad \mathrm{with}\,\, c_0=1.
\end{equation}
These relations are satisfied by 
\begin{equation}\label{eq:M_expression}
M^{ab}(\chi) = -i \left[\frac{\chi}{2} \coth{\frac{\chi}{2}}  - \frac{\chi}{2}\right]^{ab}\, .
\end{equation}
One can check explicitly that Taylor expansion of eq. \eqref{eq:M_expression} in $\chi$ reproduces all the coefficients calculated using the recursive relations in eq. \eqref{eq:recursive_relations}. 

 Once the function $M$ in Eq.\eqref{eq:M_expression} has been determined, the algebra of $\hat j^a$ and $\hat\Phi^a$ is completely defined.

We note that in order for this algebra to be consistent, the commutators have to satisfy Jacoby identity
\begin{equation}
[[\hat{\Phi}^a, \hat{j}^b], \hat{j}^c] + [[\hat{j}^b, \hat{j}^c],\hat{\Phi}^a] + [[\hat{j}^c, \hat{\Phi}^a], \hat{j}^b] =0\, ,
\end{equation}
which is equivalent to an additional constraint on $M$
\begin{equation}\label{eq:consistence_check}
[M^{ab}, \hat{j}^c] - [M^{ac}, \hat{j}^b] = igf^{bcd} M^{ad}\, .
\end{equation}
In Appendix A we verify that the Jacoby identity is in fact satisfied  at least to fourth order in expansion of eq.\eqref{eq:M_expression} in powers of $\hat\Phi$. Although we do not have a complete all order proof, one could in principle continue the order-by-order proof.  We believe that the algebra eqs.\eqref{com},\eqref{eq:M_expression} is in fact consistent and we will continue our analysis under this assumption.

We have thus found the algebra of operators $\hat\Phi^a$ and $\hat j^a$ that implements eq.\eqref{oj}. Note that expansion of $M$ in powers of $\hat\Phi$ can be recast as formal expansion of $\hat\Phi$ in powers of $\frac{\delta}{\delta\hat j^a}$. Thus to leading order we have $\hat\Phi^a=-i\frac{\delta}{\delta \hat j^a}+...$. In the dense regime where the commutators of charge densities can be neglected, our operator $\hat R$ therefore reduces precisely to the shift operator $\exp \{-\int_{\mathrm{x}_\perp}j_{\mathrm{soft}}^a(\mathrm{x}_\perp)\frac{\delta}{\delta j^a(\mathrm{x}_\perp)}\}$
used extensively in the existing literature. The previous discussion puts its use also in the dilute regime on firm mathematical basis, provided the commutation relations of $\hat\Phi^a$ are modified according to eq.\eqref{eq:M_expression}.

\subsection{The evolution}

 Having defined the charge density shift operator $\hat R$ we can write Eq.(\ref{oav}) in the form
 \begin{equation}
 {\rm Tr}[O(\hat j^a+\hat j^a_{\mathrm{soft}})\hat\rho(\Delta y)]={\rm Tr}[\hat{R}^{\dagger}O( \hat{j}^a)\hat{R} \hat\rho(\Delta y)]={\rm Tr}[O( \hat{j}^a)\hat{R} \hat\rho(\Delta y)\hat{R}^{\dagger}]
 \end{equation}
The operator $\hat R$ in this expression can be understood as acting on the density matrix $\hat\rho_v$ rather than on the observable $O$. Using this form we can define the reduced density matrix which when traced with the operator $O(\hat j)$ gives the same result as $\hat\rho(\Delta y)$ traced with $O(\hat j+\hat j_{\mathrm{soft}})$.
We thus define the evolved  CGC reduced density matrix by tracing over soft gluons
\begin{equation}\label{eq:cgc_kraus_rep}
 \hat\rho_v(\Delta y) = \mathrm{Tr}_s [\hat R\hat{\rho}(\Delta y) \hat R^\dagger]= \sum_n \langle n|\hat{R} \hat{\Omega} |0\rangle \hat\rho_v \langle 0|\hat{\Omega}^{\dagger}\hat{R}^{\dagger} |n\rangle  = \sum_n \hat{M}_n \hat\rho_v \hat{M}^{\dagger}_n
 \end{equation} 
 with $\hat{M}_n = \langle n|\hat{R} \hat{\Omega} |0\rangle$. The complete basis $\{|n\rangle \}$ represents the Fock states in the soft gluon Hilbert space. 
This procedure  technically is very similar to the standard reduction of the Hilbert space discussed in the previous section with $\hat\rho_v(\Delta y)$ playing the role of the reduced density matrix in a bipartite system.

When formulated in this way, the rapidity evolution of the CGC density matrix is formally very similar to the time evolution of the reduced density matrix of a bipartite system with the operator $\hat R\hat \Omega$ playing the role of the time evolution operator $\hat U$.

 Eq.\eqref{eq:cgc_kraus_rep} gives  the change of density matrix in the form of a Kraus representation.  As a consequence, $\hat\rho_v(\Delta y)$ has all the properties of a density matrix as long as $\hat\rho_v$ is a density matrix initially.  Note that, $\sum_n \hat{M}_n^{\dagger} \hat{M}_n =1$ as both $\hat{\Omega}$ and $\hat{R}$ are unitary operators. 
 
 \section{The Differential Form of the Evolution - the Lindblad Equation}
 
 To extract a differential equation from the Kraus representation, we need to evaluate the superoperators $\hat{M}_n$ and analyze their $\Delta y$ dependence.  The calculations of $\hat{M}_n$ can be simplified by working in the Leading Logarithmic Approximation (LLA) so that only terms that are proportional to $\alpha_s \Delta y$ on the right hand side of eq.\eqref{eq:cgc_kraus_rep} are kept. 
 \subsection{The dilute limit}
 We start by considering the dilute limit, i.e. assume that parametrically 
 $b_i^a \sim \mathcal{O}(g)$. In this regime the gluon emission operator is just the coherent operator and $\Lambda =0$.
 This is the so called KLWMIJ limit introduced in \cite{Kovner:2005nq} .
 \begin{equation}
 \Omega= \mathrm{Exp}\left\{i\int d\mathbf{x}_{\perp} b_i^a(\mathbf{x}_{\perp}) \int \frac{d\eta}{2\pi}\sqrt{2}\left(\hat{a}_i^{\dagger a}(\eta, \mathbf{x}_{\perp}) + \hat{a}_i^{a}(\eta, \mathbf{x}_{\perp})\right)\right\}
 \end{equation}
Note that we have changed the integration variable from longitudinal momentum $k^+$ to rapidity $\eta$ and an explicit numerical factor $\sqrt{2}$ follows \cite{Altinoluk:2009je}. In this limit  the dependence on $\Delta y$ becomes very transparent
\begin{equation}
\hat{M}_n =  \langle n | \hat{R}\hat{\Omega} |0\rangle=\mathrm{Exp}\left\{-\frac{\Delta y}{2\pi} b_{\alpha}b_{\alpha}\right\} \langle n| \hat{R}\, \mathrm{Exp}\left\{i\sqrt{2}b_{\alpha}\int\frac{d\eta}{2\pi} \hat{a}^{\dagger}_{\alpha}(\eta) \right\}|0\rangle \, .
\end{equation}
Note that the operator $\hat R$ has a nontrivial action on the $n$-gluon state.
 It does not change  the number of soft gluons in a Fock state but  rotates their color indices according to its definition in eq.(\ref{R})
\begin{equation}\label{eq:R_on_aa_dagger}
\begin{split}
&\hat{R} \hat{a}_i^b(k^+, \mathbf{y}_{\perp}) \hat{R}^{\dagger} = [\hat{\mathcal{R}}(\mathbf{y}_{\perp})]_{bd} \, \hat{a}^d_i(k^+, \mathbf{y}_{\perp})\, ,\\
&\hat{R} \hat{a}_i^{b\dagger}(k^+, \mathbf{y}_{\perp}) \hat{R}^{\dagger} = [\hat{\mathcal{R}}(\mathbf{y}_{\perp})]_{bd} \, \hat{a}^{d\dagger}_i(k^+, \mathbf{y}_{\perp})\,
\end{split}
\end{equation}
with 
\begin{equation}\hat{\mathcal{R}}(\mathbf{y}_{\perp}) = e^{igT^a\hat{\Phi}^a(\mathbf{y}_{\perp})} 
\end{equation}

In the LLA we need to collect terms which contribute at order $O(\alpha_s)$  to the evolution. For the virtual term we have 
\begin{equation}
\hat{M}_0 = 1  -\frac{\Delta y}{2\pi} b_{\alpha}b_{\alpha} + \mathcal{O}(g^4)
\end{equation}
and 
\begin{equation}
\hat{M}_0 \hat{\rho}_v \hat{M}^{\dagger}_0 = \hat{\rho}_v -\frac{\Delta y}{2\pi}  \left( b_{\alpha}b_{\alpha} \hat{\rho}_v + \hat{\rho}_v b_{\alpha}b_{\alpha}\right) + \mathcal{O}(g^4)\, .
\end{equation}
It is obvious that, for Fock states with even numbers of gluons, $\hat{M}_{2m}$ is at least  of order $ \mathcal{O}(g^2)$ and thus will not  contribute to the evolution at LLA. The same holds for $\hat{M}_{2m+1}$ associated with Fock states of odd numbers of gluons. The only exception is $\hat{M}_1$ related to the single gluon Fock state. For a one-gluon Fock state $|1_{\{\alpha_1, \mathbf{w}_1,\eta_1\}}\rangle = a_{\alpha_1}^{\dagger}(\mathbf{w}_1,\eta_1)|0\rangle$ with transverse position $\mathbf{w}_1$, rapidity $\eta_1$ and color index $\alpha_1$ we have,
\begin{equation}
\hat{M}_{1\{\alpha_1, \mathbf{w}_1,\eta_1\}}  =  i\sqrt{2}b_{\alpha}(\mathbf{w}_1)   \mathcal{R}_{\alpha  \alpha_1}(\mathbf{w}_1)
\end{equation}
Summing over all possible one-gluon Fock states, 
\begin{equation}
\int \frac{d\eta_1}{2\pi} \int d\mathbf{w}_{1} \,\hat{M}_{1\{\alpha_1, \mathbf{w}_1,\eta_1\}} \hat{\rho}_v \hat{M}_{1\{\alpha_1, \mathbf{w}_1,\eta_1\}}^{\dagger} = \frac{\Delta y}{\pi} \bar{b}_{\alpha} \hat{\rho}_v \bar{b}_{\alpha}
\end{equation}
with $\bar{b}_{\alpha} = \mathcal{R}^{\dagger}_{\alpha\beta}b_{\beta}$ and again we used the compact notation with index $\alpha, \beta$ representing  colors, transverse coordinates and polarizations.  The evolution equation for the density matrix follows 
\begin{equation}\label{lind}
\frac{d\hat{\rho}_v}{dy} = -\frac{1}{2\pi} (\bar{b}_{\alpha}\bar{b}_{\alpha} \hat{\rho}_v + \hat{\rho}_v \bar{b}_{\alpha}\bar{b}_{\alpha} - 2\bar{b}_{\alpha}\hat{\rho}_v \bar{b}_{\alpha} )=-\frac{1}{2\pi}\int_{x}[\bar b_a^i(x),[\bar b_a^i(x),\hat \rho_v]].
\end{equation}
In this equation we have written the virtual terms in terms of $\bar b$ rather than $b$, since the unitary operator $R$ drops out of this expression anyway.
This is the Lindblad equation for the CGC density matrix in the dilute limit. 

Eq.(\ref{lind}) is written in a somewhat convoluted form in terms of the operators $\bar b$, which contain the operator $\mathcal{R}$. It is perhaps worthwhile to make explicit the operational meaning of various factors of $\mathcal{R}$ in the right hand side of eq.(\ref{lind}).  As already mentioned, the virtual terms do not actually involve $\mathcal R$ since for unitary $\mathcal R$
\begin{equation}
\bar b_\alpha\bar b_\alpha=b_\alpha b_\alpha
\end{equation}
As for the real term, we have (suppressing the transverse coordinate)
\begin{equation}
\bar b_{\alpha}\hat{\rho}_v \bar{b}_{\alpha} =b_\gamma \mathcal{R}_{\gamma\alpha}\hat\rho_v[\hat j]\mathcal{R}^\dagger_{\alpha\beta}b_\beta=b_\gamma \left[\hat\rho_v[\hat j^a-gT^a]\right]_{\gamma\beta}b_\beta
\end{equation}
where the last term is defined by  Taylor  expanding of $\hat\rho_v$, shifting the argument $\hat j^a$ in every term by the matrix $T^a$ and finally taking the $\gamma\beta$ matrix element of the the whole expression in all products of $T$'s that arise.

In the above explicit calculation, the LLA automatically picks up terms that are linear in $\Delta y$ thus making the extraction of a differential equation from the Kraus representation straightforward. Physically indeed we can understand this from the point of view of Markovian nature of the process. The variable analogous to time in the present discussion is rapidity. Thus the requirement of short range correlations in time of the ``environment" in the CGC case translates into the requirement of short range in rapidity correlations  for the soft gluons, which are integrated over.  This is indeed the case. In the LLA the relevant ``time" scale for the change of the density matrix is $O(1/\alpha_s)$, as obvious from the differential equation eq.(\ref{lind}). The soft gluons in our approximation do not interact with each other, and thus their correlation function is free. The free propagator 
is proportional to $1/ k^+ \sim e^{-y}$, and thus the typical correlation length in rapidity space is $O(1)$. The evolution is therefore clearly in the Markovian regime which allows, at least naively speaking for the existence of differential evolution in the Lindblad form. We will come back to the discussion of Lindblad form later.

Equation eq.(\ref{lind}) may look slightly unfamiliar as it does not quite have the form of the KLWMIJ equation discussed in \cite{Kovner:2005nq} . This is because it is written for density matrix and not the weight functional $W[j]$. To get to the latter form one needs to perform an extra step, i.e. Weyl transformation. This will be the subject of the next section. But before we do that, we consider the evolution of the density matrix in the dense regime.

\subsection{The dense limit.}
As we have seen, in the limit where the hadronic wave function contains a small number of partons (the dilute limit), the Lindblad form of the evolution equation follows directly using the straightforward perturbation theory at low x. We now turn our attention to the dense limit, where we assume that the color charge density in the wave function is large, parametrically of order $1/g$. The wave function in this limit has been calculated several years ago in \cite{Altinoluk:2009je}. In this section we use the results of that paper and reinterpret them from our current point of view.

 To prepare for the calculation, note that the soft gluon emission operator  $\hat{\Omega}$, when acting on the vacuum state $|0\rangle$ can be written as
 \begin{equation}\label{ome}
\hat{\Omega}|0\rangle = \mathrm{Exp}\left\{i\sqrt{2} b_{\alpha} \int\frac{d\eta}{2\pi}[\hat{a}^{\dagger}_{\alpha}(\eta) + \hat{a}_{\alpha}(\eta)]\right\} \mathrm{Exp}\left\{-\frac{1}{2}\int\frac{d\eta}{2\pi}\frac{d\xi}{2\pi} \Lambda_{\beta\gamma}(\eta,\xi)\hat{a}^{\dagger}_{\beta}(\eta) \hat{a}^{\dagger}_{\gamma}(\xi)\right\}\mathcal{N}(\Lambda) |0\rangle
\end{equation}
Here we write out the dependence on rapidity explicitly. Other indices (color, polarization,  transverse position) are collectively represented by the Greek letters $\alpha, \beta, \gamma$.  
The matrix $\Lambda^{ab}_{ij}({\mathbf x}_\perp,{\mathbf y}_\perp,\eta_1,\eta_2)$ determines the amount of "squeezing" of the soft gluon vacuum.  As we mentioned above, it has not been calculated explicitly in \cite{Altinoluk:2009je}, however its properties relevant to the JIMWLK limit are known (see later). 
The $\mathcal{N}(\Lambda)$ is a normalization constant that only depends on $\Lambda$. Note that both $b_{\alpha}$ and $\Lambda_{\beta\gamma}$ are operators in the Hilbert space of hard gluons as they depend on the color charge density ${\mathbf j}$, and so in principle they do not commute. In eq.(\ref{ome}) all the factors of $\Lambda$ should be understood as placed to the right of $b_{\alpha}$.  In the JIMWLK limit however, where parametrically, $b=O(1/g)$ while $\Lambda=O(1)$, as was shown in \cite{Altinoluk:2009je} the order of the factors does not matter. In fact in showing that the operator $\hat \Omega$ Eq.\eqref{ome} diagonalizes the QCD Hamiltonian to leading order,  ref.\cite{Kovner:2007zu, Altinoluk:2009je} explicitly used this argument and assumed commutativity of the various factors of $b$ and $\Lambda$. We will not deviate from this assumption here and will treat these factors as commuting.

We further separate the annihilation operator $\hat{a}_{\alpha}(\eta)$ from the coherent state operator and move it to the far right acting on the vacuum state:
\begin{equation}\label{eq:omega_expression}
\begin{split}
\hat{\Omega}|0\rangle =& \mathrm{Exp}\left\{-\frac{\Delta y}{2\pi}  b_{\alpha}(1-\Lambda_{0})_{\alpha\beta} b_{\beta}\right\}\mathrm{Exp}\left\{i \sqrt{2}b_{\alpha}(1-\Lambda_{0})_{\alpha\beta} \int \frac{d\eta}{2\pi} a^{\dagger}_{\beta}(\eta)\right\}\\
&\times \mathrm{Exp}\left\{-\frac{1}{2} \int\frac{d\eta}{2\pi}\frac{d\xi}{2\pi} a^{\dagger}_{\alpha}(\eta)\Lambda_{\alpha\beta}(\eta,\xi) a^{\dagger}_{\beta}(\xi)\right\} \mathcal{N}(\Lambda)|0\rangle
\end{split}
\end{equation}
where we have defined 
 \begin{equation}
 \Lambda_{0,\alpha\beta} = \int_{-\Delta y}^{\Delta y} \frac{d\zeta}{2\pi} \Lambda_{\alpha\beta}(\zeta,\eta)
 \end{equation}
Since $\Lambda_{\alpha\beta}(\zeta,\eta)$ depends only on the rapidity difference $\zeta-\eta$ \cite{Altinoluk:2009je},
 $\Lambda_{0,\alpha\beta}$ is rapidity independent.
It does however have a nontrivial dependence on the width of the evolution step $\Delta y$. The nature of this dependence is very important. As we discussed above, we expect to have a bona fide differential evolution equation only if the correlations of the soft gluons in rapidity are short range. The function $\Lambda(\eta,\xi)$ is in fact the inverse of the correlator of the soft gluon modes. It should therefore decrease exponentially for rapidity difference greater than $\sim 1$. For such a function $\Lambda$ the dependence of $\Lambda_0$ on $\Delta y$ should be smooth with $\Lambda_0$ approximately constant for $1< \Delta y< 1/\alpha_s$. We will assume here that this is indeed the case and will treat $\Lambda_0$ as a constant independent of $\Delta y$. The results of  \cite{Altinoluk:2009je} suggest that this is valid in the JIMWLK limit, i.e. when the dense hadron scatters on a dilute target, which is the regime that concerns us  in this paper. We note that going beyond the JIMWLK limit posed some problems in  \cite{Altinoluk:2009je}, precisely for the reason that some of the soft modes in general seemed to possess long range correlations in rapidity. Our current understanding is that such long range correlations indeed are incompatible with the differential form of the evolution. It is thus possible that in order to go beyond the JIMWLK limit one would have to rethink the way in which the bipartitioning into the ``observable" system and ``environment" is done. This is however beyond the scope of the present paper.

 In eq. \eqref{eq:omega_expression}, the first exponential represents wavefunction renormalization effects that have an overall $\Delta y $ factor.  The second exponential contains the single gluon emission vertex $i b_{\alpha}(1-\Lambda_0)_{\alpha\beta}$ which is ``renormalized" relative to the dilute case by the presence of the Bogoliubov operator $B$, while the third exponential contains the double gluon emission vertex $\Lambda_{\alpha\beta}(\xi,\eta)$. 
 
  Two gluons  emitted from the same double gluon emission vertex are in general correlated in rapidity, while two gluons  emitted from two single gluon emission vertexes are uncorrelated.  
  

 We are now ready to calculate the superoperators. The fundamental difference with the dilute case, is that now not only one gluon state, but states with arbitrary number of soft gluons yield nontrivial jump operators that contribute to the evolution of the density matrix. For an $n$ soft gluon  state we have
\begin{equation}\label{eq:Mn_general_expression}
\begin{split}
\hat{M}_n = & \langle n | \hat{R}\hat{\Omega} |0\rangle=\mathrm{Exp}\left\{-\frac{\Delta y}{2\pi} b_{\alpha}b_{\beta}(1-\Lambda_0)_{\alpha\beta}\right\}  \mathcal{N}(\Lambda) \\
&\langle n| \hat{R}\, \mathrm{Exp}\left\{i\sqrt{2}b_{\alpha}(1-\Lambda_0)_{\alpha\beta}\int\frac{d\eta}{2\pi} \hat{a}^{\dagger}_{\beta}(\eta) \right\}\mathrm{Exp}\left\{- \frac{1}{2}\int\frac{d\eta}{2\pi}\frac{d\xi}{2\pi}\Lambda_{\alpha\beta}(\eta,\xi)\hat{a}^{\dagger}_{\alpha}(\eta) \hat{a}^{\dagger}_{\beta}(\xi)\right\} |0\rangle\\
\end{split}
\end{equation}
Depending on the Fock state $|n\rangle$ being considered, we separately discuss the situations when the Fock state contains zero gluons, odd number of gluons and even number of gluons.

\subsubsection{Wavefunction renormalization operator}
The superoperator $\hat{M}_0$ represents the wavefunction renormalization effects
\begin{equation}
\hat{M}_0 = \langle 0| \hat{R} \hat{\Omega} |0\rangle   = \mathrm{Exp}\left\{-\frac{\Delta y}{2\pi} b_{\alpha}b_{\beta}(1-\Lambda_0)_{\alpha\beta}\right\}  \mathcal{N}(\Lambda)\, .
\end{equation}
Up to terms  linear in $\Delta y$,
\begin{equation}
\hat{M}_0 \approx 1 -\frac{\Delta y}{2\pi}\left[b_{\alpha}b_{\beta}(1-\Lambda_0)_{\alpha\beta} \right] + \mathcal{O}(\Delta y^2)\, 
\end{equation}

Note that the wavefunction renormalization operator $\hat{M}_0$ is independent of $\hat{R}$ and we have ignored the  normalization $\mathcal{N}(\Lambda)$ factor, since it is irrelevant in the JIMWLK limit \cite{Altinoluk:2009je}. The superoperator $\hat{M}_0$  contributes to the change of density matrix through the term
\begin{equation}
\hat{M}_0 \hat{\rho}_v \hat{M}_0^{\dagger} = \hat{\rho}_v - \frac{\Delta y}{2\pi} \left[b_{\alpha}b_{\beta}(1-\Lambda_0)_{\alpha\beta} \, \hat{\rho}_v + \hat{\rho}_v\,  (1-\Lambda_0^{\dagger})_{\alpha\beta}b_{\alpha}b_{\beta} \right] + \mathcal{O}(\Delta y^2) \, .
\end{equation}

 \subsubsection{Jump operators with odd  number of gluons}
 For Fock states with odd numbers of gluons,  one needs odd number of single-gluon-emission vertices in calculating the jump operators. However, every single gluon emission brings an extra power  of $\Delta y$, since gluons produced from different single-gluon-emission vertices are uncorrelated in rapidity. The integral over rapidity of every such gluons in the amplitude and conjugate amplitude brings therefore an extra power of $\Delta y$. Thus one needs to keep only one single-gluon-emission vertex in $\hat M_{2i+1}$ in order to calculate the relevant jump operators that contribute to the differential form of the evolution equation.
 
 The explicit expression for a jump operator follows from eq. \eqref{eq:Mn_general_expression}
 \begin{equation}\label{mn}
\begin{split}
\hat{M}_n&= \langle n| \hat{R} \left(i\sqrt{2}b_{\alpha}(1-\Lambda_0)_{\alpha\beta}\int\frac{d\zeta}{2\pi} \hat{a}^{\dagger}_{\beta}(\zeta) \right) \mathrm{Exp}\left\{- \frac{1}{2}\int\frac{d\eta}{2\pi}\frac{d\xi}{2\pi}\Lambda_{\alpha\beta}(\eta,\xi)\hat{a}^{\dagger}_{\alpha}(\eta) \hat{a}^{\dagger}_{\beta}(\xi)\right\}|0\rangle\\
& =\left( i \mathcal{R}^{\dagger}_{\gamma \delta} [\sqrt{2}b(1-\Lambda_0)]_{\delta} \right) \langle n |\int\frac{d\zeta}{2\pi} \hat{a}_{\gamma}^{\dagger}(\zeta) \mathrm{Exp}\left\{- \frac{1}{2}\int\frac{d\eta}{2\pi}\frac{d\xi}{2\pi}\bar{\Lambda}_{\alpha\beta}(\eta,\xi)\hat{a}^{\dagger}_{\alpha}(\eta) \hat{a}^{\dagger}_{\beta}(\xi)\right\}|0\rangle.
\end{split}
\end{equation}
 
 Here 
 \begin{equation}\label{lambar}
 \bar{\Lambda}_{\alpha\beta} (\eta,\xi)= \mathcal{R}^{\dagger}_{\alpha\gamma} \Lambda_{\gamma\delta}(\eta,\xi) \mathcal{R}_{\delta \beta}.\end{equation}
 
 To arrive at this expression we have inserted the factor  $\hat{R}^{\dagger}\hat{R} =1$ next to the soft gluon vacuum state $|0\rangle$, used the fact that $\hat R|0\rangle=|0\rangle$ and evaluated the action of $\hat{R}$  on the soft gluon creation  and annihilation operators using Eq.\eqref{eq:R_on_aa_dagger}.

 
Importantly, the operator ordering in Eq.\eqref{mn} is such that all the operators $\mathcal{R} $ are understood  to be placed to the  left of all the factors  of the $b_{\alpha}$ and $\Lambda_{\alpha\beta}$. This follows from the fact that the operator $\hat R$ in the original expression is acting directly on the $n$-gluon state, and thus all  the factors of $\hat \Phi$ indeed are ordered to the left of all $j$-dependent factors in the original expression. Thus for example in the definition Eq.\eqref{lambar} the action of $\mathcal{R}_{\delta \beta}$ on $\Lambda$ is understood only as a color matrix rotation. This comment also applies to the rest of the formulae in this section.
  
  In the following, we explicitly calculate a few expressions of the jump operators and their action on the density matrix. This will make the dependence on $\Delta y$ more transparent.  
  
  For a one-gluon Fock state $|1_{\{\alpha_1, \mathbf{w}_1,\eta_1\}}\rangle = a_{\alpha_1}^{\dagger}(\mathbf{w}_1,\eta_1)|0\rangle$ with transverse position $\mathbf{w}_1$, rapidity $\eta_1$ and color index $\alpha_1$, the jump operator is
\begin{equation}
\hat{M}_{1\{\alpha_1, \mathbf{w}_1,\eta_1\}} = \int d^2\mathbf{z}_{1} i\sqrt{2}b_{\alpha}(\mathbf{z}_1)  [1 - \Lambda_0]_{\alpha\beta}(\mathbf{z}_1,\mathbf{w}_1)  \mathcal{R}_{\beta  \alpha_1}(\mathbf{w}_1)
\end{equation}
Note that the jump operator associated with one-gluon Fock state is independent of the rapidity index $\eta_1$. Integration over all the one-gluon Fock states produces an overall factor $\Delta y$ in the evolution of the density matrix.  The one-gluon jump operators contribute to this evolution through
\begin{equation}\label{oneg}
\begin{split}
&\hat{M}_{1}\hat{\rho}_v \hat{M}_1^{\dagger} = \sum_{\alpha_1} \int d\mathbf{w}_1\int\frac{d\eta_1}{2\pi}\hat{M}_{1\{\alpha_1, \mathbf{w}_1,\eta_1\}}  \hat{\rho}_v\hat{M}_{1\{\alpha_1, \mathbf{w}_1,\eta_1\}}^{\dagger}  \\
=&\frac{\Delta y}{\pi} \int d\mathbf{z}_1 d\mathbf{z}_2b_{\alpha}(\mathbf{z}_1) \left[\int d\mathbf{w}_1(1-\Lambda_0)_{\alpha\beta}(\mathbf{z}_1,\mathbf{w}_1) [\mathcal{R}\hat{\rho}_v\mathcal{R}^{\dagger}]_{\beta\gamma} (1-\Lambda_0^{\dagger})_{\gamma\delta}(\mathbf{w}_1, \mathbf{z}_2)\right]b_{\delta}(\mathbf{z}_2)\\
=&\frac{\Delta y}{\pi} [\bar{b}(1-\bar{\Lambda}_0]_{\alpha} \hat{\rho}_v [(1-\bar{\Lambda}_0^{\dagger})\bar{b}]_{\alpha}
\end{split}
\end{equation}
 In the last line we have reverted to the convoluted notation where  single index $\alpha$  represents the transverse position, color and polarization. Barred quantities here and below indicate the quantities that are rotated by the $\mathcal{R}$ matrix. 
 
 For a three-gluon Fock state $
|3_{\{\alpha_i, \mathbf{w}_i,\eta_i;i=1,2,3\}}\rangle = a^{\dagger}_{\alpha_1}(\mathbf{w}_1,\eta_1) a^{\dagger}_{\alpha_2}(\mathbf{w}_2,\eta_2) a^{\dagger}_{\alpha_3}(\mathbf{w}_3,\eta_3) |0\rangle $, the jump operator is
\begin{equation}
\begin{split}
&\hat{M}_{3_{\{\alpha_i, \mathbf{w}_i, \eta_i; i=1,2,3\}}} = -\int d^2\mathbf{z}_1  i\sqrt{2}b_{\alpha}(\mathbf{z}_1) \Bigg([1-\Lambda_0]_{\alpha\beta}(\mathbf{z}_1, \mathbf{w}_1) \Lambda_{\kappa\lambda}(\mathbf{w}_2,\eta_2, \mathbf{w}_3,\eta_3) \\
&+ [1-\Lambda_0]_{\alpha\kappa}(\mathbf{z}_1, \mathbf{w}_2) \Lambda_{\beta\lambda}(\mathbf{w}_1,\eta_1, \mathbf{w}_3,\eta_3) +[1-\Lambda_0]_{\alpha\lambda}(\mathbf{z}_1, \mathbf{w}_3) \Lambda_{\beta\kappa}(\mathbf{w}_1,\eta_1; \mathbf{w}_2,\eta_2) \Bigg)\\
&\times \mathcal{R}_{\beta\alpha_1}(\mathbf{w}_1)\mathcal{R}_{\kappa\alpha_2}(\mathbf{w}_2)\mathcal{R}_{\lambda\alpha_3}(\mathbf{w}_3)
\end{split}
\end{equation}
 It contains sum of all possible terms where two out of the three gluons are emitted from the same  two-gluon-emission vertex. Note that $\Lambda$ is a symmetric matrix. The contribution of the three gluon jump operator to the evolution of the density matrix is 
\begin{equation}
\hat{M}_3 \hat{\rho}_v \hat{M}_3^{\dagger} =\sum_{\alpha_1,\alpha_2,\alpha_3}\int d\mathbf{w}_1d\mathbf{w}_2d\mathbf{w}_3 \int\frac{d\eta_1}{2\pi}\frac{d\eta_2}{2\pi}\frac{d\eta_3}{2\pi}\hat{M}_{3_{\{\alpha_i, \mathbf{w}_i, \eta_i; i=1,2,3\}}}\hat{\rho}_v\hat{M}_{3_{\{\alpha_i, \mathbf{w}_i, \eta_i; i=1,2,3\}}}^{\dagger}
\end{equation}
 
 This expression has in principle nine terms.  However, for terms that involve the same two gluons connected to a two gluon  emission vertex both in $\hat M_3$ and $\hat M_3^\dagger$, integration over rapidity produces higher than linear powers in $\Delta y$. For example
 \begin{equation}
\int\frac{d\eta_1}{2\pi}\frac{d\eta_2}{2\pi}\frac{d\eta_3}{2\pi}\Lambda_{\kappa\lambda}(\mathbf{w}_{1}, \eta_{1}; \mathbf{w}_{2}, \eta_{2})\Lambda^{\dagger}_{\rho\delta}(\mathbf{w}_{1},\eta_{1};\mathbf{w}_{2},\eta_{2}) = (\Delta y)^2\int\frac{d\zeta}{2\pi}\Lambda_{\kappa\lambda}(\mathbf{w}_{1},\mathbf{w}_{2};\zeta)\Lambda^{\dagger}_{\rho\delta}(\mathbf{w}_{1},\mathbf{w}_{2};\zeta).
\end{equation}
This term therefore does not contribute to the differential  form of the evolution.

On the other hand, for the two-gluon-emission vertexes connected to different pairs of gluons, only one explicit  $\Delta y$ factor arises
\begin{equation}
\int\frac{d\eta_1}{2\pi}\frac{d\eta_2}{2\pi}\frac{d\eta_3}{2\pi}\Lambda_{\kappa\lambda}(\mathbf{w}_{1}, \eta_{1}; \mathbf{w}_{2}, \eta_{2})\Lambda^{\dagger}_{\rho\delta}(\mathbf{w}_{1},\eta_{1};\mathbf{w}_{3},\eta_{3}) = \Delta y\,  \Lambda_{0,\kappa\lambda}(\mathbf{w}_{1},  \mathbf{w}_{2})\Lambda^{\dagger}_{0,\rho\delta}(\mathbf{w}_{1},\mathbf{w}_{3})\, .
\end{equation}
These terms do contribute.

The contribution of the three gluon jump operators is thus
 \begin{equation}\label{m3}
\begin{split}
\hat{M}_3\hat{\rho}_v \hat{M}_3^{\dagger}  
=& \frac{\Delta y}{\pi} \sum_{\alpha_1, \alpha_2,\alpha_3}\int d\mathbf{w}_1d\mathbf{w}_2d\mathbf{w}_3  [\bar{b}(1-\bar{\Lambda}_0)]_{\alpha_{1}} (\mathbf{w}_{1}) \bar{\Lambda}_{0, \alpha_{2}\alpha_{3}}(\mathbf{w}_{2}, \mathbf{w}_{3})\hat{\rho}_v\\
& \quad \times  \bar{\Lambda}^{\dagger}_{0,\alpha_{1}\alpha_{2}}(\mathbf{w}_{1},\mathbf{w}_{2}) [(1-\bar{\Lambda}_0^{\dagger})\bar{b}]_{\alpha_{3}}(\mathbf{w}_{3})\\
=&\frac{\Delta y}{\pi} [\bar{b}_L(1-\bar{\Lambda}_{0,L})] \bar{\Lambda}_{0,R}^{\dagger}\bar{\Lambda}_{0,L} [(1-\bar{\Lambda}_{0,R}^{\dagger})\bar{b}_R]\hat{\rho}_v
\end{split}
\end{equation}
In the last line we have used superscripts $L$ and $R$ to indicate the position of various  factors relative to the density matrix $\hat\rho_v$, thus $\bar\Lambda_{0,L}$ indicates  that this factor $\bar\Lambda_0$ is placed to the left of $\hat\rho_v$ etc. The ordering is important as the various operators do not commute with $\hat\rho_v$.
The reason to write the expression in this particular way is that we can use convenient matrix notations, so that products in eq. (\ref{m3}) are matrix products over all indexes carried by $\bar \Lambda_0$ and $\bar b$, i.e. color, polarization and transverse coordinate.

 
 This pattern clearly generalizes to any odd $n$. Linear in $\Delta y$ contributions arise only from terms where no two gluons are emitted from the same two gluon emission vertex both in $M_n$ and $M^\dagger_n$.
Diagrammatically the terms that yield linear in $\Delta y$ contributions are depicted in Fig.1.

\begin{figure}[t]         
\centering                              
 \includegraphics[width=12cm] {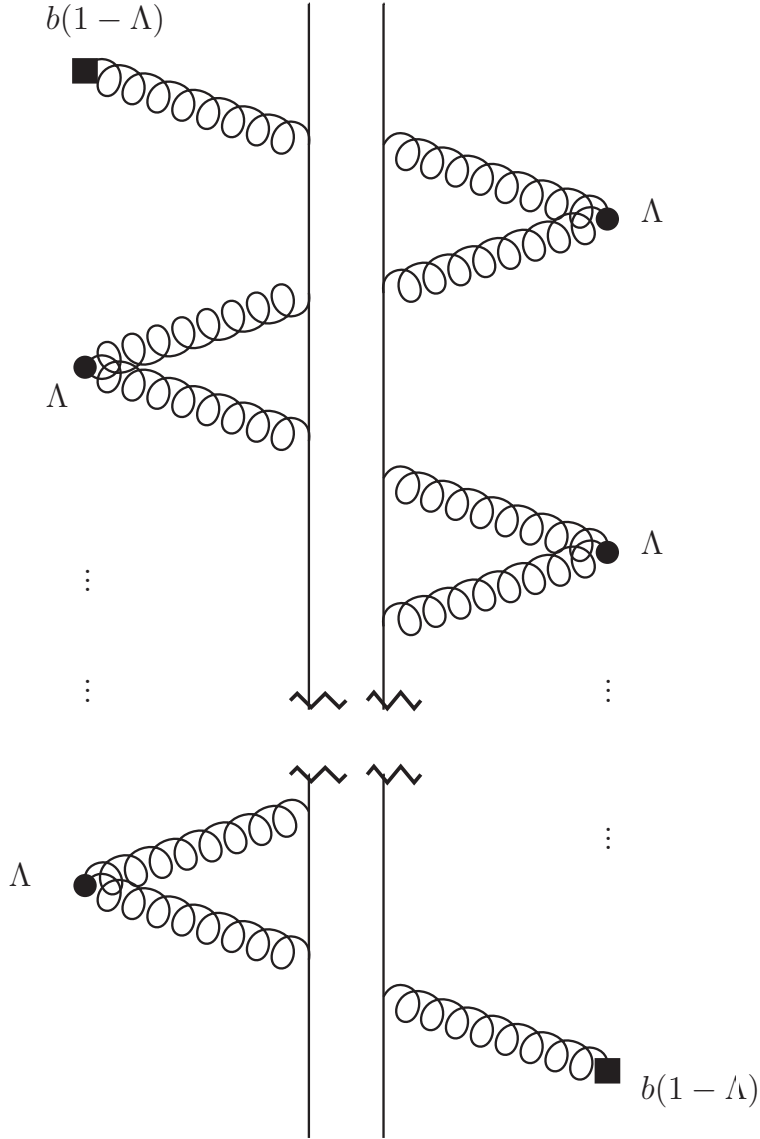}             
\caption{The diagrams involving the odd number of gluons that contribute terms linear in $\Delta y$ to the evolution of $\hat\rho$.  }
\label{f1}
\end{figure}

 Generalizing the above analysis to jump operators with $2m+1$ numbers of gluons we obtain
\begin{equation}\label{eq:jump_operator_odd_gluons}
\begin{split}
\hat{M}_{2m+1_{\{\alpha_i,\mathbf{w}_i,\eta_i; i=1,\ldots, 2m+1\}}} 
=&\sum_{P=\{i_1, \ldots, i_{2m+1}\}}[i\sqrt{2}b(1-\Lambda_0)]_{\beta_{i_1}}(\mathbf{w}_{i_1}) \prod^m_{k\neq l\neq 1} \Lambda_{\beta_{i_k}\beta_{i_l}}(\mathbf{w}_{i_k},\eta_{i_k};\mathbf{w}_{i_l},\eta_{i_l})\\
&\times \prod_{q=1}^{2m+1} \mathcal{R}_{\beta_{i_q} \alpha_{q}}(\mathbf{w}_{q})\\
\end{split}
\end{equation} 
Here $i_1, i_2, \ldots, i_{2m+1}$ is a permutation of $1, 2,\ldots, 2m+1$. The sum over  P goes over all the possible permutations. 

The action of $\hat M_{2m+1}$ on the density matrix after summing over all the possible Fock states with $2m+1$ gluons and performing the rapidity integrations, becomes
\begin{equation}
\begin{split}
&\hat{M}_{2m+1}\hat{\rho}_v \hat{M}^{\dagger}_{2m+1} = \frac{\Delta y}{\pi} [\bar{b}_L(1-\bar{\Lambda}_{0,L})](\bar{\Lambda}_{0,R}^{\dagger}\bar{\Lambda}_{0,L})^{m}[(1-\bar{\Lambda}_{0,R}^{\dagger})\bar{b}_R]\hat{\rho}_v\, .\\
\end{split}
\end{equation}
Here, just like for $\hat M_3$ the contribution comes only from those terms that do not contain a single pair of gluons emitted from the same two gluon emission vertex in $M_{2n+1}$ and $M^\dagger_{2n+1}$ as illustrated on Fig.1.

Now adding all the jump operators associated with odd numbers of gluons, their action on the density matrix is
\begin{equation}
\sum_{m=0}\hat{M}_{2m+1}\hat{\rho}_v \hat{M}^{\dagger}_{2m+1}  = \frac{\Delta y}{\pi}  [\bar{b}_L(1-\bar{\Lambda}_{0,L})](1-\bar{\Lambda}_{0,R}^{\dagger}\bar{\Lambda}_{0,L})^{-1}[(1-\bar{\Lambda}_{0,R}^{\dagger})\bar{b}_R]\hat{\rho}_v\, .
\end{equation}

  \subsubsection{Jump operators with even number of gluons}
 If the number of gluons in the Fock state is even, the gluons can either be emitted  from a two-gluon-emission vertex or from even number of single-gluon-emission vertexes.  Since gluons emitted from single-gluon-emission vertexes are uncorrelated in rapidity, the more single gluon emission vertexes are involved, the higher power of $\Delta y$ is generated. Recall that we only need to keep terms  linear in $\Delta y$.  To extract these terms, we allow either no gluons or two gluons to be emitted from the single-gluon-emission vertexes. The rest of the contributions, as we wil see are subleading in powers of $\Delta y$.
 
 The expression for the jump operators with even numbers of gluons by $\hat{M}_{2n}$ follows from eq. \eqref{eq:Mn_general_expression}
 \begin{equation}
\begin{split}
\hat{M}_{2n}= 
&\hat{M}_{2n}^{0} + \hat{M}_{2n}^{2} =\langle 2n| \hat{R}\, \mathrm{Exp}\left\{- \frac{1}{2}\int\frac{d\eta}{2\pi}\frac{d\xi}{2\pi}\Lambda_{\alpha\beta}(\eta,\xi)\hat{a}^{\dagger}_{\alpha}(\eta) \hat{a}^{\dagger}_{\beta}(\xi)\right\} |0\rangle\\
+&\langle 2n| \hat{R}\,\frac{1}{2!}\left(i\sqrt{2}b_{\gamma}(1-\Lambda_0)_{\gamma\delta}\int\frac{d\zeta}{2\pi} \hat{a}^{\dagger}_{\delta}(\zeta) \right)^2\mathrm{Exp}\left\{- \frac{1}{2}\int\frac{d\eta}{2\pi}\frac{d\xi}{2\pi}\Lambda_{\alpha\beta}(\eta,\xi)\hat{a}^{\dagger}_{\alpha}(\eta) \hat{a}^{\dagger}_{\beta}(\xi)\right\} |0\rangle\, .\\
\end{split}
\end{equation}
 We have denoted the parts with no single-gluon-emission vertex and  with two single-gluon-emission vertexes by $\hat{M}_{2n}^{0}$ and $\hat{M}^{2}_{2n}$, respectively. The action on the density matrix becomes
 \begin{equation}\label{eq:explicit_EJO_dm}
 \hat{M}_{2n}\hat{\rho}_v \hat{M}_{2n}^\dagger = \hat{M}_{2n}^0\hat{\rho}_v \hat{M}_{2n}^{0 \dagger}+ \hat{M}_{2n}^{0}\hat{\rho}_v \hat{M}_{2n}^{2\dagger}+ \hat{M}_{2n}^{2}\hat{\rho}_v \hat{M}_{2n}^{0\dagger}+ \hat{M}_{2n}^{2}\hat{\rho}_v \hat{M}_{2n}^{2\dagger}
 \end{equation}
 Just like in the case of odd number of gluons, not all the terms in eq.\eqref{eq:explicit_EJO_dm} contribute to differential evolution.
The last term in eq.\eqref{eq:explicit_EJO_dm}  contains an overall factor of $(\Delta y)^2$ and therefore can be discarded.  The first term in eq.\eqref{eq:explicit_EJO_dm}  does contain terms that are only linear in $\Delta y$, however in the dense limit it is suppressed by a power of $\alpha_s$ relative to the second and third terms, as in the dense limit $b\sim 1/g$, while $\Lambda \sim 1$.
Similar terms arise in the expansion of  the normalization factor $\mathcal{N}(\Lambda)$, which we have neglected above. We therefore discard these terms in the week coupling limit.
Only the second and third terms in eq. \eqref{eq:explicit_EJO_dm} are to be evaluated and contribute to the differential evolution of the density matrix.  

We first evaluate  $\hat{M}_{2n}^{0}$. For example, for a two-gluon Fock state $|2_{\{\alpha_1, \mathbf{w}_1,\eta_1; \alpha_2, \mathbf{w}_2,\eta_2\}}\rangle = a^{\dagger}_{\alpha_1}(\mathbf{w}_1,\eta_1)a^{\dagger}_{\alpha_2}(\mathbf{w}_2,\eta_2) |0\rangle $
 \begin{equation}
\begin{split}
 \hat{M}_{2_{\{\alpha_1,\mathbf{w}_1,\eta_1; \alpha_2,\mathbf{w}_2,\eta_2\}}}^{0} 
 & = - \Lambda_{\kappa\lambda}(\mathbf{w}_1,\eta_1;\mathbf{w}_2,\eta_2) \mathcal{R}_{\kappa\alpha}(\mathbf{w}_1) \mathcal{R}_{\lambda\alpha_2}(\mathbf{w}_2)= -\bar{\Lambda}_{\alpha_1\alpha_2}(\mathbf{w}_1,\eta_1;\mathbf{w}_2,\eta_2)\\
 \end{split}
\end{equation}
Generalization to Fock states with $2m$ gluons is straightforward
\begin{equation}
\begin{split}
\hat{M}_{2m_{\{\alpha_i,\mathbf{w}_i,\eta_i; i=1,\ldots, 2m\}}}^{0} 
=&\left(-\frac{1}{2}\right)^m \frac{1}{m!} \sum_{P=\{i_1, \ldots, i_{2m}\}} \prod_{k=1}^m \bar{\Lambda}_{\alpha_{i_{2k-1}}\alpha_{i_{2k}}}(\mathbf{w}_{i_{2k-1}},\eta_{i_{2k-1}};\mathbf{w}_{i_{2k}},\eta_{i_{2k}}) \\
\end{split}
\end{equation}
 The summation is over all the permutations. The prefactor $(1/2)^m$ is needed to account for the fact that $\Lambda$  is a symmetric matrix. The factor $1/m!$  takes care of the fact that two permutations that differ only by ordering of some  pairs of indices and nothing else, give identical contributions which only need to be counted once. 
 
 To calculate $\hat{M}_{2n}^{2}$ we start with the simple situation of two-gluon Fock state
 \begin{equation}
\begin{split}
\hat{M}_{2_{\{\alpha_1,\mathbf{w}_1,\eta_1; \alpha_2,\mathbf{w}_2,\eta_2\}}}^{2} 
&=- \int d^2\mathbf{z}_1 d^2\mathbf{z}_3 b_{\alpha}(\mathbf{z}_1) b_{\rho}(\mathbf{z}_3)\bigg( [1-\Lambda_0]_{\alpha\beta}(\mathbf{z}_1, \mathbf{w}_1)[1-\Lambda_0]_{\rho\delta}(\mathbf{z}_3, \mathbf{w}_2)\bigg)\\
&\quad \times \mathcal{R}_{\beta\alpha_1}(\mathbf{w}_1) \mathcal{R}_{\delta\alpha_2}(\mathbf{w}_2)\\
&=-[\bar{b}(1-\bar{\Lambda}_0)]_{\alpha_1}(\mathbf{w}_1) [\bar{b}(1-\bar{\Lambda}_0)]_{\alpha_2}(\mathbf{w}_2)
\end{split}
\end{equation}
 Generalization to  Fock states with $2m$ gluons leads to
\begin{equation}
\begin{split}
\hat{M}^{2}_{2m_{\{\alpha_i,\mathbf{w}_i,\eta_i;i=1,\ldots, 2m\}}}&=-\frac{1}{(m-1)!}\left(-\frac{1}{2}\right)^{m-1} \sum_{P=\{i_1,\ldots, i_{2m}\}} [\bar{b}(1-\bar{\Lambda}_0)]_{\alpha_{i_1}}(\mathbf{w}_{i_1}) [\bar{b}(1-\bar{\Lambda}_0)]_{\alpha_{i_2}}( \mathbf{w}_{i_2})\\
&\quad\times\prod_{k=2}^{m}\bar{\Lambda}_{\alpha_{i_{2k-1}}\alpha_{i_{2k}}}(\mathbf{w}_{i_{2k-1}},\eta_{i_{2k-1}}, \mathbf{w}_{i_{2k}},\eta_{i_{2k}})\,.\\
\end{split}
\end{equation}
 The  action of the two gluon jump operator on the density matrix is 
\begin{equation}
\begin{split}
\hat{M}_2^{2}\hat{\rho}_v \hat{M}_2^{0\dagger}& = \sum_{\alpha_1\alpha_2}\int d\mathbf{w}_1d\mathbf{w}_2\frac{d\eta_1}{2\pi}\frac{d\eta_2}{2\pi}[\bar{b}(1-\bar{\Lambda}_0)]_{\alpha_1}(\mathbf{w}_1) [\bar{b}(1-\bar{\Lambda}_0)]_{\alpha_2}(\mathbf{w}_2) \hat{\rho}_v \bar{\Lambda}^{\dagger}_{\alpha_1\alpha_2}(\mathbf{w}_1,\eta_1;\mathbf{w}_2,\eta_2)\\
&=\frac{\Delta y}{2\pi} [\bar{b}_L(1-\bar{\Lambda}_{0,L})]\bar{\Lambda}_{0,R}^{\dagger}[\bar{b}_L(1-\bar{\Lambda}_{0,L})] \hat{\rho}_v\, .
\end{split}
\end{equation}
For the four gluon operator we similarly find
\begin{equation}
\hat{M}_4^{2}\hat{\rho}_v \hat{M}_4^{ 0\dagger} = \frac{\Delta y}{2\pi}  [\bar{b}_L(1-\bar{\Lambda}_{0,L})]\bar{\Lambda}_{0,R}^{\dagger}\bar{\Lambda}_{0,L} \bar{\Lambda}_{0,R}^{\dagger}[\bar{b}_L(1-\bar{\Lambda}_{0,L})] \hat{\rho}_v\, .
\end{equation}
Summing up all the jump operators with even numbers of gluons, we get
\begin{equation}
\sum_{m=1}^{\infty}\hat{M}_{2m}^{2}\hat{\rho}_v \hat{M}_{2m}^{0\dagger} =\frac{\Delta y}{2\pi}  [\bar{b}_L(1-\bar{\Lambda}_{0,L})](1-\bar{\Lambda}_{0,R}^{\dagger}\bar{\Lambda}_{0,L})^{-1} \bar{\Lambda}_{0,R}^{\dagger}[\bar{b}_L(1-\bar{\Lambda}_{0,L})] \hat{\rho}_v\, .
\end{equation}
 and its complex conjugate
 \begin{equation}
 \sum_{m=1}^{\infty}\hat{M}_{2m}^{0}\hat{\rho}_v \hat{M}_{2m}^{2\dagger} =\frac{\Delta y}{2\pi}  [\bar{b}_R(1-\bar{\Lambda}_{0,R}^{\dagger})]\bar{\Lambda}_{0,L}(1-\bar{\Lambda}_{0,R}^{\dagger}\bar{\Lambda}_{0,L})^{-1}[\bar{b}_R(1-\bar{\Lambda}_{0,R}^{\dagger})] \hat{\rho}_v\, .
 \end{equation}
 \subsubsection{All together now.}
We now put together the above results. The wave function renormalization operator and the jump operators with even numbers of gluons contribute to what one might call ``the virtual part" of the evolution:
\begin{equation}
\begin{split}
&\hat{M}_0\hat{\rho}_v \hat{M}^{\dagger}_0 +\sum_{n=1}^\infty\hat{M}_{2n} \hat{\rho}_v \hat{M}^{ \dagger}_{2n}\\
= &\hat{\rho}_v- \left(\frac{\Delta y}{2\pi}\right)\left[ \bar{b}_L(1-\bar{\Lambda}_{0,L})(1-\bar{\Lambda}^{\dagger}_{0,R}\bar{\Lambda}_{0,L})^{-1} (1-\bar{\Lambda}^{\dagger}_{0,R})\bar{b}_L\right] \, \hat{\rho}_v\\
&- \left(\frac{\Delta y}{2\pi} \right)\left[\bar{b}_R(1-\bar{\Lambda}_{0,L})(1-\bar{\Lambda}^{\dagger}_{0,R}\bar{\Lambda}_{0,L})^{-1}  (1-\bar{\Lambda}^{\dagger}_{0,R})\bar{b}_R\right]\,\hat{\rho}_v\, .
\end{split}
\end{equation}
Adding contributions from jump operators with odd numbers of gluons we get
\begin{equation}\label{eq:deltaY_changes}
\begin{split}
&\hat{M}_0\hat{\rho}_v \hat{M}^{\dagger}_0 +\sum_{n=1}^\infty\hat{M}_{2n} \hat{\rho}_v \hat{M}^{ \dagger}_{2n} + \sum_{n=0}^\infty\hat{M}_{2n+1} \hat{\rho}_v \hat{M}^{ \dagger}_{2n+1}\\
=&\hat{\rho}_v -\left(\frac{\Delta y}{2\pi}\right)\left[(\bar{b}_L -\bar{b}_R)(1-\bar{\Lambda}_{0,R}^{\dagger})(1- \bar{\Lambda}_{0,L}\bar{\Lambda}_{0,R}^{\dagger})^{-1}(1-\bar{\Lambda}_{0,L}) (\bar{b}_L -\bar{b}_R)\right]\hat{\rho}_v\, .
\end{split}
\end{equation}
\subsubsection{Operator ordering.}
 As we have mentioned earlier, the relative ordering of  operators of $b$ and $\Lambda_0$ in Eq.\eqref{eq:deltaY_changes} is not important. It is however important to keep track of the ordering of the classical fields $b_L$ and $b_R$ relative to the density matrix. More precisely, in the JIMWLK limit one can change the order of various  factors in Eg.\eqref{eq:deltaY_changes} as long as each factor $(\bar b_L-\bar b_R)$ is  kept as a unit and is commuted with any other operator in question..
 
  
The argument for that was given in \cite{Altinoluk:2009je} , and we reproduce it here  for completeness.
 
 Recall that the JIMWLK limit is obtained when parametrically $b\sim O(1/g)$ and $\Lambda_0\sim O(1)$, and additionally the evolution equation should be expanded to order $\alpha_s$. The latter expansion gives the leading contribution when the dense hadron scatters on a dilute target.
 
 Since both $b$ and $\Lambda_0$ are functions of $j$, the commutator between $b$ and $\Lambda_0$ can be estimated as
 \begin{equation}
 (b\Lambda_0 - \Lambda_0 b) = \frac{\delta b}{\delta j^{a}} [j^a, j^b] \frac{\delta \Lambda_0}{\delta j^b} = igf^{abc} \frac{\delta b}{\delta j^{a}} j^c \frac{\delta \Lambda_0}{\delta j^b} \sim \mathcal{O}(g)\, .
 \end{equation} 
 The difference $\bar{b}_L -\bar{b}_R$ can also be estimated as 
 \begin{equation}
 (b_L -b_R)\hat\rho_v =  \left(b\hat{\rho}_v - \hat{\rho}_v b\right) = \frac{\delta b}{\delta j^a} [j^a, j^b] \frac{\delta \hat{\rho}_v}{\delta j^b} = igf^{abc}   \frac{\delta b}{\delta j^a} j^c \frac{\delta \hat{\rho}_v}{\delta j^b} \sim \mathcal{O}(g)\hat\rho_v
 \end{equation}
 and barred quantities are color rotated by $\hat{\mathcal{R}}\sim \mathcal{O}(1)$.The difference between $\Lambda_{0,L}$ and $\Lambda_{0,R}$ is estimated similarly
 \begin{equation}
( \Lambda_{0,L} -\Lambda_{0,R})\hat\rho_v = \Lambda_0 \hat{\rho}_v -\hat{\rho}_v \Lambda_0 = \frac{\delta \Lambda_0}{\delta j^a} [j^a , j^b] \frac{\delta \hat{\rho}_v}{\delta j^b} = igf^{abc} \frac{\delta \Lambda_0}{\delta j^a} j^c \frac{\delta \hat{\rho}_v}{\delta j^b}\sim \mathcal{O}(g^2)\hat\rho_v\, .
 \end{equation}
 Since the factor $(\bar{b}_L -\bar{b}_R)^2$  on the right hand side of eq. \eqref{eq:deltaY_changes} is already of order of $\alpha_s$, the ordering between $b$ and $\Lambda_0$ and the difference between $\Lambda_{0,L}$ and $\Lambda_{0,R}$ contribute to higher orders in $\alpha_s$ and therefore can be ignored in the JIMWLK limit as long as one does not order the operators differently in the terms containing $b_L$ and $b_R$. Thus for example, one can substitute in Eq.\eqref{eq:deltaY_changes} 
 \begin{equation}(1-\bar{\Lambda}_{0,R}^{\dagger})(1- \bar{\Lambda}_{0,L}\bar{\Lambda}_{0,R}^{\dagger})^{-1}(1-\bar{\Lambda}_{0,L})\rightarrow )(1-\bar{\Lambda}_{0,R}^{\dagger})(1- \bar{\Lambda}_{0,R}\bar{\Lambda}_{0,R}^{\dagger})^{-1}(1-\bar{\Lambda}_{0,R})
 \end{equation}or 
 \begin{equation}(1-\bar{\Lambda}_{0,R}^{\dagger})(1- \bar{\Lambda}_{0,L}\bar{\Lambda}_{0,R}^{\dagger})^{-1}(1-\bar{\Lambda}_{0,L})\rightarrow)(1-\bar{\Lambda}_{0,L}^{\dagger})(1- \bar{\Lambda}_{0,L}\bar{\Lambda}_{0,L}^{\dagger})^{-1}(1-\bar{\Lambda}_{0,L})\end{equation} as a whole, without breaking the factor $(\bar{b}_L -\bar{b}_R)^2$ into separate pieces.
 
 \subsubsection{The Lindblad form, finally.}
 We now simplify eq.\eqref{eq:deltaY_changes} using the results of \cite{Altinoluk:2009je}. 
 First we note that the function of $\bar\Lambda$ appearing in eq.\eqref{eq:deltaY_changes}  can be represented as a square if we indeed forget about the difference between $\Lambda_L$ and $\Lambda_R$. Define formally
 \begin{equation}
 \Theta=\sqrt{(1-\Lambda\Lambda^\dagger)^{-1}}
 \end{equation}
 We can then write
 \begin{equation}
 (1-\bar{\Lambda}^{\dagger})(1- \bar{\Lambda}\bar{\Lambda}^{\dagger})^{-1}(1-\bar{\Lambda})=\bar N^\dagger\bar N
 \end{equation}with
 \begin{equation}\label{n}
 N=\Theta(1-\Lambda)
 \end{equation}
 In fact the matrix $\Theta$ appears naturally in the calculation of \cite{Altinoluk:2009je}.
 Since the soft gluon vacuum is a squeezed state due to the presence of the Bogoliubov operator ${\cal\hat B}$, it is the Fock space vacuum of the Bogoliubov transformed set of creation and annihilation operators, related to the original gluon operators $a^\dagger$ and $a$ by
 \begin{equation} \hat{\beta}_{\rho} = \Theta_{\rho\sigma } \hat{a}_{\sigma} + \Phi_{\rho\sigma} \hat{a}^{\dagger}_{\sigma}; \ \ \ \ \ \ \ \ \ \ \ 
\hat{\beta}_{\rho}^{\dagger} = \Theta^{\ast}_{\rho\sigma} \hat{a}^{\dagger}_{\sigma} + \Phi^{\ast}_{\rho\sigma} \hat{a}_{\sigma} 
\end{equation}
where $\Theta$ and $\Phi$ are constrained by the unitarity condition
\begin{equation}
\Theta\Theta^{\dagger} -\Phi\Phi^{\dagger}=1
\end{equation}
The following relation was also derived in \cite{Altinoluk:2009je}
 \begin{equation}
 \Lambda = \Theta^{-1} \Phi
 \end{equation}
 Using these relations we find
\begin{equation}\label{eq:multiparticle_correction}
(1-\bar{\Lambda}^{\dagger})(1- \bar{\Lambda}\bar{\Lambda}^{\dagger})^{-1}(1-\bar{\Lambda})
=(\bar{\Theta}^{\dagger} - \bar{\Phi}^{\dagger}) (\bar{\Theta} -\bar{\Phi}).
\end{equation}
with the usual definitions $\bar{\Theta} = \mathcal{R}^{\dagger}\Theta \mathcal{R}$ and $\bar{\Phi} = \mathcal{R}^{\dagger} \Phi\mathcal{R}$.   This is indeed the same as eq.\eqref{n} with
\begin{equation}
N=\Theta-\Phi
\end{equation}

After integration over rapidities, eq. \eqref{eq:multiparticle_correction} becomes 
\begin{equation} \label{eq:N_perp_to_Lambda}
\frac{\Delta y}{2\pi}(1-\bar{\Lambda}_{0}^{\dagger})(1- \bar{\Lambda}_{0}\bar{\Lambda}_{0}^{\dagger})^{-1}(1-\bar{\Lambda}_{0}) = \frac{\Delta y}{2\pi} \bar{N}_{\perp}^{\dagger} \bar{N}_{\perp}\, ,
\end{equation}
where $N_{\perp} = \int d\eta (\Theta -\Phi)$  has been calculated in \cite{Altinoluk:2009je}
\begin{equation}\label{eq:Nperp_exp}
N_{\perp} = [1-l-L] = \left[\delta_{ij} - \partial_i \frac{1}{\partial^2}\partial_j -D_i\frac{1}{D^2} D_j\right]^{ab} (\mathbf{x}_{\perp}, \mathbf{y}_{\perp})\, .
\end{equation}
where the covariant derivative is defined as $D_i^{ab} = \delta^{ab}\partial_i -igT^{e}_{ba}b_i^e$.

Substituting eq.\eqref{eq:N_perp_to_Lambda} into eq. \eqref{eq:deltaY_changes} , we obtain
\begin{equation}\label{eq:classical_approximated_equation}
\hat{\rho}_v(\Delta y) = \hat{\rho}_v - \frac{\Delta y}{2\pi} (\bar{b}_L -\bar{b}_R) \bar{N}_{\perp}^{\dagger} \bar{N}_{\perp} (\bar{b}_L - \bar{b}_R)\hat{\rho}_v\, .
\end{equation}
To write this explicitly as an operator equation we need to make the choice of whether to place the factor $\bar N^\dagger \bar N$ to the right or to the left of the density matrix, as both choices, as well as some others are equivalent in the JIMWLK limit. Additionally we need to specify the ordering between the operators $b$ and $\Lambda$ on one hand and factors of $R$ on another, which has been scrambled in the calculation above. It is not our goal here to carefully restore the correct ordering, as only the JIMWLK limit of this expression is strictly speaking under control. We therefore simply choose the specific ordering which reproduces JIMWLK as well as gives an evolution equation which preserves the Hermiticity of the density matrix also away from the JIMWLK limit.

We take 
\begin{equation}
\bar{N}^{\dagger}_{\perp} \bar{N}_{\perp} = \mathcal{R}^{\dagger} (N^{\dagger}_{\perp}N_{\perp})\mathcal{R}, \qquad \bar{b} = b\mathcal{R}\, , \, \ \ \ \bar b^\dagger =\mathcal R^\dagger b
\end{equation}
where the operator ordering is now specified. With these definitions we write the evolution as

\begin{equation}\label{eq:nonlindblad_eq_cgc}
\begin{split}
\frac{d\hat{\rho}_v}{dy} =& -\frac{1}{4\pi} \left(\hat{\rho}_v \bar{b}_{\alpha} (\bar{N}^{\dagger}_{\perp} \bar{N}_{\perp})_{\alpha\beta} \bar{b}^{\dagger}_{\beta} + \bar{b}^{\dagger}_{\beta}\bar{b}_{\alpha} \hat{\rho}_v (\bar{N}^{\dagger}_{\perp} \bar{N}_{\perp})_{\alpha\beta}  - \bar{b}_{\alpha} \hat{\rho}_v (\bar{N}^{\dagger}_{\perp} \bar{N}_{\perp})_{\alpha\beta} \bar{b}_{\beta}^{\dagger} - \bar{b}^{\dagger}_{\beta} \hat{\rho}_v \bar{b}_{\alpha} (\bar{N}^{\dagger}_{\perp} \bar{N}_{\perp})_{\alpha\beta} \right)+h.c.\\
=&-\frac{1}{4\pi} \left[\bar{b}^{\dagger}_{\beta}, \left[\bar{b}_{\alpha}, \hat{\rho}_v\right](\bar{N}^{\dagger}_{\perp} \bar{N}_{\perp})_{\alpha\beta} \right]+h.c.\\
\end{split}
\end{equation}


 Written out explicitly
\begin{equation}\label{eq:nonlindblad_eq_cgc_explicit}
\begin{split}
\frac{d\hat{\rho}_v}{dy} =& -\frac{1}{4\pi} \Big[\hat{\rho}_v b_{\alpha} (N^{\dagger}_{\perp} N_{\perp})_{\alpha\beta} b_{\beta} + \mathcal{R}^{\dagger}_{\alpha\omega}b_{\omega}b_{\rho}\mathcal{R}_{\rho\beta} \hat{\rho}_v \mathcal{R}^{\dagger}_{\beta\lambda}(N^{\dagger}_{\perp} N_{\perp})_{\lambda\kappa} \mathcal{R}_{\kappa\alpha}\\
& - b_{\rho}\mathcal{R}_{\rho\alpha} \hat{\rho}_v \mathcal{R}^{\dagger}_{\alpha\omega}(N^{\dagger}_{\perp} N_{\perp})_{\omega\beta}b_{\beta} - \mathcal{R}^{\dagger}_{\alpha\omega}b_{\omega} \hat{\rho}_v b_{\beta} (N^{\dagger}_{\perp} N_{\perp})_{\beta\kappa}\mathcal{R}_{\kappa\alpha} \Big]+h.c.\, .\\
\end{split}
\end{equation}

Several words on the nature of the evolution of $\hat\rho_v$ as given by \eqref{eq:nonlindblad_eq_cgc_explicit}.   As we discussed earlier, the only relevant characteristic of a state in the valence Hilbert space is its representation of the color $SU(N)$ (at each spatial point). Therefore the valence Hilbert space on which $\hat\rho_v$ is defined is a direct sum of all the possible subspaces labelled by different representations of $SU(N)$, i.e. the values of all the Casimir operators at each point $\mathcal{J}= \{\hat{j}^a(\mathbf{x}_{\perp} )\hat{j}^a(\mathbf{x}_{\perp}), \ d^{abc}\hat{j}^a(\mathbf{x}_{\perp} )\hat{j}^b(\mathbf{x}_{\perp})\hat{j}^c(\mathbf{x}_{\perp})\ ...\}$, so that $\mathcal{H} = \oplus \mathcal{H}_{\mathcal{J}}$. Since the density matrix itself depends only on $j^a$, it is a block diagonal operator on this Hilbert space and has nonvanishing matrix elements only between states that belong to the same representation ${\mathcal J}$.  
The same is true for the operators  $b_{\alpha}$ and $N_{\alpha\beta}$ since they also are functions of $j^a$ only. Thus if not for the operator ${\mathcal R}$ in Eq.\eqref{eq:nonlindblad_eq_cgc_explicit} the evolution would mix the matrix elements of $\hat\rho_v$ in a given representation ${\mathcal J}$ only between themselves.
 The presence of  the operator $\hat{\mathcal{R}}$ changes the nature of the evolution. It shifts $j^a$ to $j^a + gT^a$ and therefore mixes matrix elements of $\hat\rho_v$ in one representation with those in another representation with an additional adjoint added in. 
 The operator $\hat{\mathcal{R}}$ is thus the only source of communication between subspaces of different $\mathcal{J}$ in the evolution. 
 Note that even though such cross talk between different representations exists throughout the evolution, the matrix $\hat\rho_v$ remains block diagonal if it was chosen to be block diagonal at the initial rapidity, since for such an initial condition the right hand side of eq.\eqref{eq:nonlindblad_eq_cgc_explicit} is a function of $j^a$.

Note that  Eq.\eqref{eq:nonlindblad_eq_cgc_explicit}  is somewhat more general than the JIMWLK equation. To obtain the original JIMWLK equation (apart from invoking quantum-classical correspondence which is the subject of the next section) one has to expand \eqref{eq:nonlindblad_eq_cgc_explicit} to second order in 
 in $\Phi^a$. 
 In fact eq.\eqref{eq:nonlindblad_eq_cgc_explicit} as written here contains both, the JIMWLK limit when expanded in $\Phi^a$ as well as the KLWMIJ limit when expanded in $j^a$. It can therefore be viewed as an interpolating form of the evolution equation for density matrix between the dense and dilute regimes, just like the corresponding equation for the (quasi)probability density functional $W[{\mathbf j}]$ in \cite{Altinoluk:2009je}.

The expansion to leading order in $\hat\Phi^a$ can be performed directly in Eq.\eqref{eq:nonlindblad_eq_cgc_explicit}. One has to be careful however, since apart from expanding the explicit dependence on $\hat\Phi^a$ in operators ${\mathcal R}$ one also needs to expand the commutators of $\hat\rho$ and $b$. This is easier done in a somewhat roundabout way, namely transforming the operator equation into the equation for the quasi probability function, performing the expansion there, and then returning to the operator equation using the Wigner - Weyl transformation. We will do precisely this in the next section after introducing the quasiclassical correspondence  between the quantum dynamics and dynamics on classical phase space. Here we only present the result of this exercise. The evolution equation in the JIMWLK limit turns out to be
\begin{equation}\label{qlin}
\frac{d\hat{\rho}_v}{dy}=-\frac{1}{2\pi} \int d^2\mathbf{z}_{\perp} [\hat Q_i^{a}(\mathbf{z}_{\perp}),[ \hat Q_i^a(\mathbf{z}_{\perp}),\hat\rho_v]]
\end{equation}
where
\begin{equation}\label{eq:BL_BR1}
\hat Q_i^a({\mathbf x})=\int_{{\mathbf z}}\left[U({\mathbf x})\left(D_i\frac{1}{D^2} - \partial_i\frac{1}{\partial^2}\right)D\partial\right]^{ab}(\mathbf{x},\mathbf{z}) \hat\Phi^b({\mathbf z})\, .
\end{equation}
and the matrix $U$ is defined as
\begin{equation}\label{u}
U(\mathbf{x}_{\perp}) = \mathcal{P} \mathrm{exp}\left[ ig\int_{\mathcal{C}} d^2\mathbf{y}_{\perp} \cdot b^c(\mathbf{y}_{\perp})T^c \right]
\end{equation}
with the path $\mathcal{C}$ starting from infinity on the transverse plane and ending at some point $\mathbf{x}_{\perp}$\footnote{The most common form of the eikonal scatttering amplitude one finds in the literature is a lightlike Wilson line. This is the right definition in a gauge which has a nonvanishing light cone component of the vector potential $A^-$. Our discussion here is set in the lightcone gauge in which $A^-=0$. In this gauge the scattering amplitude is given by the transverse Wilson line at $x^+\rightarrow\infty$, which is defined in Eq.\eqref{u}.}.

We note that this is precisely the equation for density matrix proposed in \cite{Armesto:2019mna}.

A comment is in order on the form of the evolution equation. First we note that Eq.\eqref{qlin} is of the Lindblad type. Thus  we find that both in the dilute and dense limits the CGC density matrix satisfies Lindblad type equations. Interestingly however, our interpolating equation Eq.\eqref{eq:nonlindblad_eq_cgc},\eqref{eq:nonlindblad_eq_cgc_explicit} does not have the Lindblad form.  One might wonder if this absence of Lindblad form is simply an artifact of our approximation. After all, the calculations that lead to Eq.\eqref{eq:nonlindblad_eq_cgc} are under full control only in the two limits. However the reason for deviation from Lindblad in rapidity evolution seems to be very general, and it rather looks like very special conditions have to be satisfied in order for Lindblad form to hold.


We drew an analogy from open quantum systems by treating the valence  gluons as ``the system" and the soft guons as ``the environment".  However, time evolution and rapidity evolution are very different concepts. In the former situation, the system degrees of freedom and the environment degrees of freedom are well specified from the beginning and do not change over time. The interaction between the system and the environment is assumed to be Markovian which holds if the system degrees of freedom are slow while the environment degrees of freedom are fast.  Time correlations of the environment degrees of freedom are assumed to be local in time compared to the long time scale on which the changes of the system occur, and this leads to Lindblad form of the evolution equation via Eq.\eqref{amp}. 

In the case of rapidity evolution, however, the separation between the ``environment" - the fast soft gluonic degrees of freedom and the ``system" - the slow valence partons is not fixed, but instead the separation boundary moves together with the evolution parameter. As the rapidity increases one therefore integrates over additional degrees of freedom, namely those whose rapidity label is between the old and new values of the evolution parameter - $\Delta y$.  The rapidity thus appears not only as a parameter of the evolution analogous to time, but also as the label of the quantum states which are being integrated out in the process. This integration over additional degrees of freedom is part and parcel of rapidity evolution, and  the increment in the density matrix $\Delta\hat\rho_v$  proportional to $\Delta y$ arises due to this integration. Thus the "Markovian" regime (i.e. short correlation length of the environment modes  in rapidity) albeit sufficient to guarantee existence of differential evolution, does not guarantee that this evolution is of Lindblad form. In fact it is easy to trace that the additional integration over the rapidity label is in fact the reason why the general argument that leads to Lindblad form for time evolution in quantum mechanics, is violated in our calculation. 

The crucial missing piece is Eq.\eqref{amp}. As discussed in Section II, in an evolving quantum system for small $\Delta t$ the probability $\hat M^\dagger_n(dt)\hat M_n(dt)\propto dt$ is proportional to $dt$ {\it for every environment state $n$} save the vacuum. One can then define the jump operator $\hat L_n$ via $\hat M_n(dt)=\sqrt{dt}\hat L_n$ and the Lindblad form follows. On the other hand, in the case of rapidity evolution the factor $\Delta y$ arises only as a result of the integration over rapidity label of the gluon states in the rapidity window $[y\ ; \ y+\Delta y]$. As a result we have  
\begin{equation}\int_{\eta=y}^{y+\Delta y} \hat M^\dagger_n(\eta,\Delta y)\hat M_n(\eta,\Delta y)\propto \Delta y
\end{equation}
but the probabilities for individual states with fixed $\eta$ do not scale with $\Delta y$. It thus does not appear to be possible in general to define a jump operator unless $\hat M_n(\eta)$ has very special properties. 
Indeed, examining our calculation, for example  in Eq.\eqref{eq:explicit_EJO_dm}, we realize that the fact that only the terms $\hat M^0_{2n}\hat\rho_v\hat M^{2\dagger}_{2n}+\hat M^2_{2n}\hat\rho_v\hat M^{0\dagger}_{2n}$ contribute at linear order in $\Delta y$  is precisely due to the integration over the rapidities of the $2n$ gluons. This is also the reason this contribution cannot be written in the standard Lindblad form $\Delta y \hat L_{2n}\hat\rho_v\hat L^\dagger_{2n}$. The same is evidently true also for the odd gluon contributions. 

Nevertheless in some special cases the Lindblad form may be attainable.
For example, if $\hat M(\eta,\Delta y)=\hat M(\Delta y)$, i.e. if the probability of a particular state does not depend on the rapidity of the gluon, the jump operator can indeed be defined. This is precisely the situation we encounter in the derivation in the dilute (KLWMIJ) limit. In this case the coherent operator ${\mathcal C}$ involves only the gluon creation operator integrated over rapidity and as a result the probability $\hat M^\dagger_nM_n$ does not depend on the rapidity label of the gluons in the state $n$. This then allows to take the "square root" of the probability and define the corresponding jump operator $\hat L_n$ which ensures that evolution is in Lindblad form. It is more difficult to trace the origin of the Lindblad form in the dense limit. However, given that JIMWLK and KLWMIJ limits are dual to each other, it is not surprising that such a form indeed exists.


In this section we have discussed the energy evolution in terms of the CGC density matrix. This is not the way it has been formulated in the literature so far. In the next section we show how to relate the two formulations.

\section{From the Lindblad Equation to the JIMWLK Equation via Quantum-Classical Correspondence}
In the previous section we have derived the rapidity evolution equation for the CGC density matrix. To turn this into the conventional JIMWLK evolution equation we will invoke a variant of the Quantum-Classical correspondence, which for simple quantum mechanical systems has been studied 50 years ago, for review see \cite{Hillery:1983ms}. To start with, we present this analysis as it appears in  \cite{Hillery:1983ms, Cahill:1969iq, Agarwal:1971wc, Agarwal:1971wb}. 

\subsection{Quantum mechanics in phase space}

For simplicity, consider a system with one degree of freedom equipped with the canonical variables $\hat{q}$ and $\hat{p}$ that satisfy the cannonical commutation realtion $[\hat{q}, \hat{p}] = i$. The state of the system is described by the density matrix operator $\hat{\rho}$. Observables are expressed as functions of $\hat{q}, \hat{p}$, e.g. $\hat{A}\equiv A[\hat{p}, \hat{q}]$.  

Suppose we want to represent a calculation of quantum expectation values in a form similar to averaging over classical distribution in phase space, i.e.
\begin{equation}
\mathrm{Tr}(\hat{\rho}\hat{A})  = \int dp dq {\mathcal A}[q, p] W[q,p]
\end{equation}
This can be achieved if one can find a one to one correspondence between an arbitrary quantum operator $ A(\hat q,\hat p)$ and a corresponding classical function ${\mathcal A}(q,p)$, and additionally similar correspondence for the density matrix $\hat\rho\rightarrow W(q,p)$.

In principle one can devise different mappings that achieve this goal. 
One widely used mapping is the Wigner-Weyl transformation that maps fully symmetrized quantum operators in the Hilbert space to the corresponding classical functions in the phase space (and back). The familiar form of the Wigner transformation uses eigenstates of the $\hat{q}$ operator and defines the classical phase space functions via
\begin{equation}
W[q,p] = \int dz e^{ipz} \langle q-\frac{z}{2} |\hat{\rho}| q + \frac{z}{2}\rangle 
\end{equation}
and 
\begin{equation}\label{eq:wigner_transformation}
A_w[q,p] =\int dz e^{ipz} \langle q-\frac{z}{2} |A[\hat{q},\hat{p}]| q + \frac{z}{2}\rangle \, .
\end{equation}
The Wigner function $W[q,p]$ that corresponds classically to the density matrix is often called the quasi probability distribution function on the phase space. To formulate the mapping in a basis-independent way, one follows Weyl's correspondence rule which associates fully symmetrized operators in Hilbert space to classical functions in phase space.  

Consider the following representation of an  operator $G(\hat p,\hat q)$
\begin{equation}\label{eq:Gpq_fourier_transform}
G[\hat{p}, \hat{q}] = \int f(u, v) e^{i (u\hat{p} + v\hat{q})} du dv\, .
\end{equation}
This can be regarded as an operator Fourier transformation. 
First of all, note that the operator $G$ written in this form is necessarily symmetric under permutations of $\hat q$ and $\hat p$. This is straightforward to see by expanding the exponential in Taylor series. This is however not a restriction on the set of operators one can consider, as {\it any} operator function can be written in a symmetric form utilizing the commutation relation between $\hat p$ and $\hat q$. The simplest example of such symmetrization is $\hat p\hat q=\frac{1}{2}(\hat p\hat q+\hat q\hat p)-\frac{i}{2}$. One can easily convince oneself that any polynomial of $\hat p$ and $\hat q$ can be written in a symmetric form of this type.

Given the representation eq.\eqref{eq:Gpq_fourier_transform} we define a classical function on the phase space via
\begin{equation}\label{eq: Fpq_fourier_transform}
F(p,q) = \int f(u, v) e^{i(up + vq)} du dv\, .
\end{equation}
This is Weyl's rule for correspondence between quantum operators and classical functions on phase space.
Note that under Weyl's rule, the same Fourier kernel $f(u, v)$ is used in eqs \eqref{eq: Fpq_fourier_transform} and \eqref{eq:Gpq_fourier_transform}.  

From Eqs. \eqref{eq: Fpq_fourier_transform} and \eqref{eq:Gpq_fourier_transform},  the mapping between $F(p,q)$ and $G[\hat{p}, \hat{q}]$ can be represented as
\begin{equation}\label{eq:basis_independent_weyl_mapping}
\begin{split}
& G[\hat{p},\hat{q}] = \int \frac{dp}{2\pi}\frac{dq}{2\pi} F(p,q) \Delta_w(p-\hat{p}, q-\hat{q})\, ,\\
& F(p,q) = \mathrm{Tr} \big( G[\hat{p},\hat{q}]\Delta_w(p-\hat{p}, q-\hat{q})\big)\\
\end{split}
\end{equation}
with 
\begin{equation}
\Delta_w(p-\hat{p}, q-\hat{q}) = \int du dv e^{ -i [u(p-\hat{p}) + v(q-\hat{q})]}\, .
\end{equation}
The Weyl mapping kernel $\Delta_w(p-\hat{p}, q-\hat{q})$ is a functional of both canonical operators $\hat{p},\hat{q}$ and phase space classical variables $q, p$.  It has the following properties
\begin{equation}
\Delta_w(p-\hat{p}, q-\hat{q}) =\Delta^\dagger_w(p-\hat{p}, q-\hat{q}) ; 
\ \ \ \ \ \ \mathrm{Tr} [\Delta_w(p-\hat{p}, q-\hat{q}) ] =1
\end{equation}
\begin{equation}
\mathrm{Tr} [\Delta_w(p-\hat{p}, q-\hat{q}) \Delta_w^{\dagger}(p^{\prime}-\hat{p}, q^{\prime}-\hat{q}) ] = (2\pi)^2\delta(p-p^{\prime})\delta(q-q^{\prime}). 
\end{equation}


The basis independent mapping in eqs. \eqref{eq:basis_independent_weyl_mapping} reproduces the familiar Wigner transformation eq. \eqref{eq:wigner_transformation} when the eigenbasis of $\hat{q}$ is used. 
\begin{equation}
\begin{split}
F(p,q) 
&=\int\frac{dq^{\prime}}{2\pi} \langle q^{\prime}| G[\hat{p},\hat{q}]\Delta_w(p-\hat{p}, q-\hat{q})|q^{\prime}\rangle\\
& = \int\frac{dq^{\prime}}{2\pi}\int du dv e^{ -i (up+vq)}e^{iv(q^{\prime}-\frac{1}{2}u)} \langle q^{\prime}|G[\hat{p},\hat{q}]|q^{\prime}-u\rangle \\
& =\int\frac{dq^{\prime}}{2\pi}\int du e^{-iup} 2\pi\delta(q-q^{\prime} + \frac{1}{2}u) \langle q^{\prime}|G[\hat{p},\hat{q}]|q^{\prime}-u\rangle \\
& = \int du e^{-iup} \langle q+\frac{1}{2}u|G[\hat{p},\hat{q}] |q-\frac{1}{2} u\rangle 
\end{split}
\end{equation}
where we have used the Baker-Campbell-Hausdorff formula, and 
\begin{equation}e^{iu\hat{p}}|q^{\prime}\rangle = |q^{\prime} -u\rangle; \ \ \ \ \ \ 
e^{i(u\hat{p} + v\hat{q})}|q^{\prime}\rangle  = e^{iu\hat{p}} e^{iv\hat{q}} e^{-\frac{i}{2} uv} |q^{\prime}\rangle = e^{ivq^{\prime}} e^{-\frac{i}{2} uv} |q^{\prime} -u\rangle\, .
\end{equation}
Thus the Weyl's correspondence rule provides a basis independent mapping between quantum operators in Hilbert space and classical functions in phase space. \\

Using eq. \eqref{eq:basis_independent_weyl_mapping} one finds that  expectation values of quantum observables can be calculated as  weighted integrals in the phase space 
\begin{equation}
\mathrm{Tr}(\hat{\rho}\hat{A})  = \int dp dq A_w[q, p] W[q,p]
\end{equation}
with
\begin{equation}
W(p,q)=\mathrm{Tr} \big( \hat \rho\Delta_w(p-\hat{p}, q-\hat{q})\big)
\end{equation}

Note that once the operator  $A_s[\hat{q}, \hat{p}]$ is written in a fully symmetrized form with respect to $\hat{q}, \hat{p}$, its associated Wigner-Weyl mapped classical function can be obtained by simply replacing the quantum operators $\hat{q},\hat{p}$ with their classical counterparts $q, p$, $A_s[\hat{q}, \hat{p}] \rightarrow  A_s[q, p] = A_w[q,p]$.  

Consider now a product of two operators $\hat{F}[\hat{q},\hat{p}] = \hat{A}[\hat{q}, \hat{p}] \hat{B}[\hat{q}, \hat{p}]$. Even if  both $\hat{A}[\hat{q},\hat{p}]$ and $\hat{B}[\hat{q},\hat{p}]$ are fully symmetrized, their product as a function of $\hat p$ and $\hat q$ does not necessarily have a fully symmetrized form, and therefore the Wigner-Weyl transform of a product is not equal to product of two Wigner-Weyl transforms. Instead, the correct procedure to obtain the Wigner-Weyl transformation for a product of two observables is 
\begin{equation}
F_w[q,p] = A_w[q, p] e^{\frac{\Lambda}{2i}} B_w[q,p] = B_w[q,p] e^{-\frac{\Lambda}{2i}} A_w[q,p]
\end{equation}
with $\Lambda = \frac{\overleftarrow{\partial}}{\partial p} \frac{\overrightarrow{\partial}}{\partial q}-\frac{\overleftarrow{\partial}}{\partial q} \frac{\overrightarrow{\partial}}{\partial p}$, where  the derivatives act on functions on the left or on the right as indicated by the arrows.  

An alternative representation for the transformation of products of operators can be achieved by introducing the left and right Bopp operators
\begin{equation}\label{boppl}
Q_L = q + \frac{i}{2}\frac{\partial}{\partial p}\, ,\quad P_L = p - \frac{i}{2}\frac{\partial}{\partial q}
\end{equation}
and 
\begin{equation}\label{boppr}
Q_R = q - \frac{i}{2}\frac{\partial}{\partial p}\, , \quad P_R = p + \frac{i}{2}\frac{\partial}{\partial q}\, .
\end{equation}
The two sets of Bopp operators labelled by ``L'' and ``R'' are operators in phase space rather than  Hilbert space, as they act on classical functions of $p$ and $q$. Note that the pairs $(Q_L,P_L)$ and $(Q_R,-P_R)$ as operators in phase space form the same Heisenberg algebra as do $(\hat q,\hat p)$ in the Hilbert space.
Using Bopp operators, the Wigner-Weyl transformation of a product of two observables is expressed as
\begin{equation}\label{fqp}
F_w[q,p] = A_s[Q_L, P_L] B_w[q,p] = B_s[Q_R,P_R]A_w[q,p]\, .
\end{equation}
Here $A_s[Q_L, P_L]$ is obtained by replacing $\hat{q}$ and $\hat{p}$ in $ A_s[\hat{q}, \hat{p}]$ by $Q_L$ and $P_L$, respectively, and similarly for $B[Q_R, P_R]$.  One can use either  set of Bopp operators, depending on whether one uses them in the left factor or the right factor of the product. The two expressions in eq.(\ref{fqp}) are equivalent. 
 The action of Bopp operators represents the additional symmetrization rearrangement necessary in order to represent a product of two symmetrized operators in a completely symmetrized form. 


As an application of the Wigner-Weyl transformation formalizm, consider the equation of motion for classical quasi distribution $W$ that follows for the quantum Liouville equation for the density matrix
\begin{equation}\label{liouville}
\frac{d\hat{\rho}}{dt}  = -i [\hat{H},\hat{\rho}] 
\end{equation}
with the Hamiltonian $\hat{H}\equiv H[\hat{q},\hat{p}]$.  Performing the Wigner-Weyl transformation of eq.(\ref{liouville}) we obtain
\begin{equation}
\begin{split}
\frac{d W[q,p]}{dt} &= -i (H_w[q,p] e^{\frac{\Lambda}{2i}} W[q,p]  - W[q,p] e^{\frac{\Lambda}{2i}} H_w[q,p])\\
& = 2 H_w[q,p] \sin{\left(\frac{\Lambda}{2i}\right)} W[q,p]
\end{split}
\end{equation}
or equivalently using the Bopp operators
\begin{equation}
\frac{d W[q,p]}{dt} = -i\left(H[Q_L, P_L] - H[Q_R, P_R]\right) W[q,p]\, .
\end{equation}
On the other hand, for a fully symmetrized observable $A_s[\hat{q}, \hat{p}]$, the Heisenberg equation is 
\begin{equation}
\frac{dA_s[\hat{q}, \hat{p}]}{dt} = i [\hat{H}, A_s[\hat{q}, \hat{p}]]\, .
\end{equation}
Its phase space formulation becomes
\begin{equation}
\begin{split}
\frac{dA_s[q,p]}{dt} &= i (H(Q_L, P_L) - H(Q_R, P_R)) A_s[q, p]\\
&= \left\{ i [\hat{H}, A_s[\hat{q}, \hat{p}]]\right\}_{s}\bigg|_{\hat{q}, \hat{p}\rightarrow q, p}
\end{split}
\end{equation}
In the second line, symmetrization for $H[\hat{q},\hat{p}] A_s[\hat{q}, \hat{p}]$ and $A_s[\hat{q}, \hat{p}]H[\hat{q},\hat{p}] $ with respect to $\hat{q},\hat{p}$ is performed before replacing $\hat{q}, \hat{p}$ with $q, p$, respectively.

\subsection{Quantum-classical correspondence for $SU(N)$ charges}
In the context of the JIMWLK evolution, the relation between the ``probability density functional" $W[\mathbf{j}]$ and the density matrix $\hat\rho$ is similar to that between $W[p,q]$ and $\hat\rho$ in quantum mechanics as described in the previous subsection. In this subsection we make this relation explicit. Much of this discussion already appears in the literature, e.g. in \cite{Kovner:2005uw}, but the relation to the classical-quantum correspondence and Wigner-Weyl transformation has not been elucidated in the past.

Unlike the canonical case discussed above, we are now dealing with the system whose  phase space is spanned by the generators of the $SU(N)$ group $j^a(\mathbf{x}_{\perp})$. It is important to stress that  the components of color charge density are coordinates on the phase space, and not on configurations space. The Hilbert space of the corresponding quantum system is spanned by the quantum operators $\hat{j}^a(\mathbf{x}_{\perp})$ that satisfy the $SU(N)$ commutation relations $[\hat{j}^a(\mathbf{x}_{\perp}, \hat{j}^b(\mathbf{y}_{\perp})] = igf^{abc}\hat{j}^c(\mathbf{x}_{\perp})\delta(\mathbf{x}_{\perp} - \mathbf{y}_{\perp})$.  Note that although the full Hilbert space of CGC requires introduction of the operators $\hat\Phi^a({\mathbf x}_\perp)$, the observables that are currently considered in all calculations are only functions of the color charge density $\hat j^a({\mathbf x}_\perp)$. It is thus sufficient for our purposes to discuss the quantum-classical correspondence  for operators that depend only on ${\mathbf j}({\mathbf x}_\perp)$. One must keep in mind however that if one wishes to generalize the framework along the lines of \cite{Armesto:2019mna}, this correspondence has to be extended to include also functions of $\hat\Phi({\mathbf x}_\perp)$.

  For quantum systems of spins, most notably the $SU(2)$ group,  the mapping between the Hilbert space and the phase space has long been studied \cite{Stratonovich:1956,Varilly:1989sv, Brif:1997km, Brif:1998pw}.  These studies mostly rely on introducing a particular (over)complete basis (ususally generalized coherent states) and working in a  fixed  representation of the underlying Lie group. Our situation is slightly different, since as discussed above the valence Hilbert space is a direct sum of different representations of $SU(N)$. We therefore cannot fix the representation and instead  will rely on the operator properties of the quantum-classical correspondence. The purpose of this section is, drawing analogy to the canonical case to provide the mapping between quantum operators in valence Hilbert space and classical variables in non-Abelian phase space, as well as relation between the quantum density matrix and ``classical"
quasi probability distribution.

We concentrate on operators in the Hilbert space which can be written as functions of $\hat j^a(\mathbf{x})$ and that are fully symmetric with respect to interchange of the different color components of $\hat{j}^a$. If an operator is not written in a fully symmetrized form, it can always be brought into such form by repeated use of the basic commutation relations of $\hat j^a$.  This has been explicitly demonstrated in \cite{Kovner:2005uw}. 

We will construct the analog of Weyl's quantum-classical correspondence where the classical counterpart of a quantum operator is obtained by replacing an operator $\hat{j}^a$ with its classical counterpart $ j^a$ once a quantum operator $ \mathcal{\hat O}$ is expressed in terms of fully symmetrized products of $\hat j^{a_i}$, so that  $ \mathcal{\hat O}=\mathcal{O}_s(\hat{j}^a)$ , where the subscript ``s'' indicates a fully symmetric function. 
This is the most straightforward generalization of  Weyl's quantum-classical correspondence.  In the following the spatial coordinates of $\hat{j}^a(\mathbf{x}_{\perp})$ are suppressed since $\hat{j}^a(\mathbf{x}_{\perp})$ with different transverse coordinates commute. All the nontrivial action therefore happens at the same transverse coordinate. 


In analogy with the discussion of the Weyl's correspondence rules for canonical operators in the previous subsection, we adopt the following rules for correspondence between an operator and a classical function on the phase space
\begin{equation}\label{eq:weyl_rule_color}
F[\mathbf{j}] = \int  f(\mathbf{\alpha}) e^{\,i \mathbf{\alpha}\cdot \mathbf{j}} d\mathbf{\alpha}\, ,\quad G_{s}[\hat{\mathbf{j}}] = \int  f(\mathbf{\alpha}) e^{\,i \mathbf{\alpha}\cdot\hat{ \mathbf{j}}} d\mathbf{\alpha}\, .
\end{equation}
Note that for $SU(N_c)$, the color index runs from $1$ to $N_c^2-1$. The above Fourier transformations are understood as $N_c^2-1$-variate transformations. 
It is easy to see by Taylor expanding the second of eq.\eqref{eq:weyl_rule_color} that $G_s$ is a fully symmetric function of $\hat j^a$.


 Note that eq.\eqref{eq:weyl_rule_color} is an operator relation, and is not limited to any particular representation of the $SU(N)$ group, but is rather valid on all the valence Hilbert space.

Eq. \eqref{eq:weyl_rule_color} leads to the following relation between $G_s[\hat{\mathbf{j}}]$ and  $F[\mathbf{j}]$ 
\begin{equation}\label{eq:operators_to_functions}
G_s[\hat{\mathbf{j}}] = \int\, d\mathbf{j}\, F[\mathbf{j}] \Delta_{W} (\mathbf{j} ,\hat{\mathbf{j}}) 
\end{equation}
with the mapping kernel 
\begin{equation}
\Delta_{W} (\mathbf{j} , \hat{\mathbf{j}})  = \int d\mathbf{\alpha} \, e^{-i \mathbf{\alpha} \cdot \mathbf{j}}e^{ i\mathbf{\alpha}\cdot \hat{\mathbf{j}}}\, .
\end{equation}
Our definition is such that the integration $\int d\mathbf{j}$ is over all real valued $j^a$ with a simple integration measure on $R^{N^2-1}$. 

The $\Delta_W(\mathbf{j}, \hat{\mathbf{j}})$ maps classical functions $F[\mathbf{j}]$ to quantum operators $G_s[\hat{j}^a]$. 
By requiring that for a fully symmetrized operator $G_s[\hat{\mathbf{j}}]$ the corresponding classical function in phase space is just $G_s[\mathbf{j}]$, so that  $F[\mathbf{j}] = G_s[\mathbf{j}]$ we can also find the inverse mapping. 

Let us write this mapping in the suggestive form:
\begin{equation}\label{eq:functions_to_operators}
G_s[\mathbf{j}] = \mathrm{Tr} \left(G_s[\hat{\mathbf{j}}] \tilde{\Delta}_W(\mathbf{j}, \hat{\mathbf{j}})\right)\, .
\end{equation}
Substituting eq. \eqref{eq:operators_to_functions} into eq. \eqref{eq:functions_to_operators}, one obtains the condition that $\tilde{\Delta}_W[\mathbf{j}, \hat{\mathbf{j}}]$ must satisfy
\begin{equation}\label{eq:kernel_conditions}
\mathrm{Tr}\left(\Delta_W[\mathbf{j}_1, \hat{\mathbf{j}}]\tilde{\Delta}_W[\mathbf{j}_2, \hat{\mathbf{j}}]\right) = \delta(\mathbf{j}_1-\mathbf{j}_2)\,.
\end{equation}
The following expression of $\tilde{\Delta}_W[\mathbf{j}, \hat{\mathbf{j}}]$ solves this constraint:
\begin{equation}\label{eq:deltatilde}
\tilde{\Delta}_W[\mathbf{j}, \hat{\mathbf{j}}] = \int dg_{\alpha}\,  e^{i\alpha\cdot \mathbf{j}} \, d_r(\hat{\mathbf{j}}) \, e^{i\alpha \cdot \hat{\mathbf{j}}}
\end{equation}
Here  $dg_{\alpha}$ is the Haar measure over the $SU(N)$ group. Each group element $g_{\alpha}$ is labelled by the parameters $\alpha$. The factor $d_r(\hat{\mathbf{j}})$ denotes the dimension of the particular representation $r$ and is  viewed here as a function of $\hat{\mathbf{j}}$ which depends only on the  Casimir operators  of the Lie algebra. 
 The function is such that for a given representation its numerical value is equal to the dimension of this representation. 
 
 To prove that $\tilde{\Delta}_W[\mathbf{j}, \hat{\mathbf{j}}] $ satisfies eq. \eqref{eq:functions_to_operators}, we use the Peter-Weyl theorem \cite{Barut:1986} for representations of Lie group 
\begin{equation}
\sum_{r=1}^{\infty} \sum_{j,k=1}^{d_r} d_r D^{(r)T}_{jk}(g_{\alpha_1})D^{(r)}_{jk}(g_{\alpha_2}) = \delta(g_{\alpha_1} - g_{\alpha_2}).
\end{equation}
Here $D^{(r)}(g_{\alpha})$ is the representation of group element $g_{\alpha}$ and $r$ indicates all the irreducible, inequivalent representations. As above, $d_r$ is the dimension of the representation $r$. 

Using eq.\eqref{eq:deltatilde} in the left hand side of eq.\eqref{eq:kernel_conditions}, and remembering that summation over all representations $r$ is a part of  tracing over the valence Hilbert space, we recover the right hand side of eq.\eqref{eq:kernel_conditions}.

The above expression is the formal definition of the kernel $\tilde\Delta$, however for all practical purposes one does not need to know its explicit form. This is because  the classical counterpart of the operator $ G[\hat{\mathbf  j}]$ is simply obtained by substitution $\hat{\mathbf j}\rightarrow {\mathbf j}$ once the operator $\hat G$ is written in the fully symmetrized form.

Now we can establish the relation between quantum average of operators in Hilbert space and phase space weighted integrations
\begin{equation}
\mathrm{Tr}(\hat{\rho}G_s[\hat{\mathbf{j}}] ) = \int\, d\mathbf{j} \, G_s[\mathbf{j}] \,  \mathrm{Tr}(\hat{\rho} \Delta_{W} (\mathbf{j}, \hat{\mathbf{j}}) ) =  \int\, d\mathbf{j} \, G_s[\mathbf{j}] \, W[\mathbf{j}]\, .
\end{equation}
with the  classical weight functional
\begin{equation}\label{eq:density_matrix_to_wigner_function}
W[\mathbf{j}] = \mathrm{Tr}\left(\hat{\rho} \Delta_{W} (\mathbf{j}, \hat{\mathbf{j}})\right)
\end{equation}
One can check that $\int d\mathbf{j} \, W[\mathbf{j}] =1$ using $\mathrm{Tr} \hat{\rho} =1$ and  so the classical weight function $W[\mathbf{j}]$ has the interpretation of quasi probability distribution.  

The mapping back from the classical weight functional to the density matrix is through the kernel $\tilde{\Delta}_W[\mathbf{j}, \hat{\mathbf{j}}]$. Again, one does not need an explicit form of $\tilde\Delta$ to perform this mapping. The practical way to do it, is to expand $W[\mathbf j]$ in Taylor series, and then substitute in every term
\begin{equation}
j^{a_1}...j^{a_n}\rightarrow \frac{1}{n!}\sum_{i_1,...,i_n}\hat j^{a_{i_1}}...\hat j^{a_{i_n}}
\end{equation}
where the summation goes over all possible permutations of $(1,...,n)$.

The other issue we need to understand in order to formulate evolution in the classical phase space approach is how to extend the mapping for products of quantum operators. 
In principle, this involves generalizing Moyal's star-product to a general Lie algebra. 
However, rather than taking this general mathematical approach, the particular realization for $SU(N)$ algebra has been worked out in  \cite{Kovner:2005uw, Altinoluk:2013rua}.
Consider a product of two operators with each one written in the symmetrized form
\begin{equation}
 G[\hat{\mathbf{j}}]=A_s[\hat{\mathbf{j}}]B_s[\hat{\mathbf{j}}]
\end{equation}
The analog of the transformation eq.\eqref{fqp} for the present case is
\begin{equation}
G_s[{\mathbf{j}}]=A_s[{\mathbf{j}}_L]B_s[{\mathbf{j}}]=B_s[{\mathbf{j}}_R]A_s[{\mathbf{j}}]
\end{equation}
where the appropriate Bopp operators are defined as
\begin{equation}
\begin{split}
j_L^a =&j^b\left[\frac{\tau}{2}\coth{\frac{\tau}{2}} + \frac{\tau}{2}\right]^{ba}\\
= &j^a + \frac{1}{2} j^b \left(gT^e\frac{\delta}{\delta j^e}\right)_{ba}+\frac{1}{12} j^b \left(gT^e\frac{\delta}{\delta j^e}\right)^2_{ba} -\frac{1}{720}j^b\left(gT^e\frac{\delta}{\delta j^e}\right)^4_{ba} +\ldots\\
j_R^a  =&j^b\left[\frac{\tau}{2}\coth{\frac{\tau}{2}} - \frac{\tau}{2}\right]^{ba}\\
 =& j^a - \frac{1}{2} j^b \left(gT^e\frac{\delta}{\delta j^e} \right)_{ba}+\frac{1}{12} j^b \left(gT^e\frac{\delta}{\delta j^e}\right)^2_{ba} -\frac{1}{720}j^b\left(gT^e\frac{\delta}{\delta j^e}\right)^4_{ba}+ \ldots\\
\end{split}
\end{equation}
with 
\begin{equation}
\tau = gT^e\frac{\delta}{\delta j^e}
\end{equation}

The Bopp operators $j^a_L$ and $j^b_R$ act on functions in the phase space rather than the Hilbert space. 
 It is straightforward if somewhat tedious to explicitly check that, similarly to Bopp operators defined in Eqs.\eqref{boppl},\eqref{boppr}, the $SU(N)$ phase space Bopp operators   $j^a_L$ and $-j^a_R$ form the same $SU(N)$ algebra as the operators $\hat{j}^a$ on the Hilbert space.  In addition, $j^a_L$ and $j^b_R$ commute $[j^a_L, j^b_R] =0$.  
 
 We find it interesting to note that the functional form of the  Bopp operators $j^a_R$ involves exactly the same function as in eq.\eqref{eq:M_expression}, which ensured correct operator properties of the charge shift operator $\hat R$.
 
 As a corollary to this discussion consider Hermitian conjugation in Hilbert space
  \begin{equation}
 (\hat A\hat B)^\dagger=\hat B^\dagger \hat A^\dagger
 \end{equation}
 As discussed above the classical correspondence is
 \begin{equation}
 \hat A\hat B\rightarrow A_s(J_L)B_s(j); \ \ \ \ \ \ \hat B^\dagger\hat A^\dagger\rightarrow B^*(j)A^*(J_R)
 \end{equation}
 Thus the Hermitian conjugation operation is represented by complex conjugation in conjunction with changing left (right) Bopp operators int right (left) Bopp operators
 \begin{equation}\label{herm}
 (... )^\dagger\rightarrow (L\leftrightarrow R)^*
 \end{equation}
 
 \subsection{The evolution equation for the quasi probability distribution.}
 Using the correspondence rules described above we can now rewrite the evolution equations eq. \eqref{eq:nonlindblad_eq_cgc},\eqref{eq:nonlindblad_eq_cgc_explicit} for the density matrix as the evolution equation for the quasi probability distribution $W[{\mathbf{j}}]$.
 
 The right hand side of eq.\eqref{eq:nonlindblad_eq_cgc_explicit} contains product of operators $(N_{\perp}^{\dagger}N_{\perp})$, $b$, $\mathcal{R}$ and $\hat\rho$. Performing Wigner-Weyl transformation the density matrix $\hat\rho$ becomes $W[j^a(\mathbf{x})]$. The operator $b$  becomes $b_L^{\alpha}=b^{\alpha}[j_L^a(\mathbf{x}_{\perp}); \mathbf{x}_{\perp}]$ or $b_R^{\alpha}=b^{\alpha}[j_R^a(\mathbf{x}_{\perp}); \mathbf{x}_{\perp}]$, depending on its position relative to the factor $\hat\rho$ in eq.\eqref{eq:nonlindblad_eq_cgc_explicit}. The operator $N_{\perp}^{\dagger}N_{\perp}$ becomes $N_{\perp,R}^{\dagger}N_{\perp,R}$ with $N_{\perp, R} \equiv N_{\perp}[j^a_R(\mathbf{x}_{\perp})]$.

  
  Additionally we  need to understand  how the operator $\hat{\mathcal{R}}$ is mapped to the phase space. To do this, we note that for a fully symmetrized operator $\hat{O}_s[\hat{j^a}]$, the action of $\hat{\mathcal{R}}$ operator is
  \begin{equation} \hat{\mathcal{R}}\hat{O}_s[j^a] \hat{\mathcal{R}} ^{\dagger}= \hat{O}_s[\hat{j}^a +gT^a] \rightarrow \mathcal{R}_p O_s[j^a]
  \end{equation}
   with the phase space shift operator 
   \begin{equation}\mathcal{R}_p = e^{gT^a\frac{\delta}{\delta j^a}}
   \end{equation}
Therefore, we should simply replace the action of the operator $\hat{\mathcal{R}}$ with the phase space shift operator $\mathcal{R}_p$.  

It is now straightforward to write the equation for $W$. We obtain

\begin{equation}\label{eq:phase_space_lindblad}
\frac{d W[j^a]}{dy}  = -\frac{1}{4\pi} \Bigg[(\tilde{b}_L^{\alpha} - b_R^{\alpha})^{\dagger}(N^{\dagger}_{\perp,R}N_{\perp,R} )_{\alpha\beta}(\tilde{b}_L^{\beta} -b_R^{\beta}) +h.c.\Bigg]W[j^a] 
\end{equation}
with $\tilde{b}_L^{\alpha} = b_L^{\beta}\mathcal{R}_{p}^{\beta\alpha}$ and hermitian conjugation defined in Eq.\eqref{herm}.  
Eq.\eqref{eq:phase_space_lindblad} is the final form of the evolution equation in the classical phase space formulation.
When $j^a$ are considered as coordinates on a classical phase space, this equation is interpretable as a Focker-Planck equation for the quasi probability phase space distribution $W$.

 One can now take various limits to reproduce the results known in the literature. In particular, assuming that ${\mathbf{j}}$ is small and expanding the right hand side of eq.\eqref{eq:phase_space_lindblad} to second order in ${\mathbf{j}}$ one straightforwardly recovers the so called KLWMIJ equation \cite{Kovner:2005nq},\cite{Kovner:2005en}.

 Alternatively, keeping all orders in ${\mathbf{j}}$, but expanding to second order in $\delta/\delta{\mathbf{j}}$ one reproduces the JIMWLK equation. 
 This last expansion is a little more involved, but it is performed explicitly in \cite{Kovner:2007zu}. We reproduce the derivation here for completeness.  
To reproduce the JIMWLK kernel, we truncate $\mathcal{R}_p$ to first order in $\delta/\delta j^a$ and expand $b_L$ and $b_R$ around $b(j^a)$. We only need to keep first order terms, since Eq.\eqref{eq:phase_space_lindblad} contains a factor $(\tilde b_L-b_R)^2$.

At this order there is no need to expand $N_{\perp,R}$ or $N_{\perp,L}$ so that both are substituted by $N_\perp (j)$.
We have
\begin{equation}
\begin{split}
&b_L\mathcal{R}_p - b_R  = b_i^a[j_L] \mathcal{R}^{ab}_p - b_i^b[j_R] \\
\simeq & \left(b_i^a[j] + \frac{\delta b_i^a}{\delta j^e}(j_L^e -j^e)\right)\left(\delta_{ab}+ gT^d_{ab}\frac{\delta}{\delta j^d}\right) - \left(b_i^b[j] +\frac{\delta b_i^b}{\delta j^e} (j_R^e -j^e)\right)\\
\simeq&gb_i^aT^d_{ab}\frac{\delta}{\delta j^d} + \frac{\delta b_i^b}{\delta j^e}g j^cT^d_{ce}\frac{\delta}{\delta j^d}\\
=&gb_i^a(\mathbf{x}_{\perp})T^d_{ab}\frac{\delta}{\delta j^d(\mathbf{x}_{\perp})} + \int d^2\mathbf{z}_{\perp}\frac{\delta b_i^b(\mathbf{x}_{\perp})}{\delta j^e(\mathbf{z}_{\perp})} gj^c(\mathbf{z}_{\perp})T^d_{ce}\frac{\delta}{\delta j^d(\mathbf{x}_{\perp})}\\
=&i\left[\partial_i - D_i\frac{1}{\partial D} D\partial \right]^{bd}(\mathbf{x},\mathbf{z})\, \frac{\delta}{\delta j^d(\mathbf{z}_{\perp})}
\end{split}
\end{equation}

In the last line we have used $igT^{e}_{ba}b_i^e = \delta^{ab}\partial_i - D_i^{ab} $ 
and $
igT^e_{ba}j^e = -(\partial D - D\partial)^{ab} = igj^eT^a_{eb} $ as well as $
\frac{\delta b_i^b(\mathbf{x}_{\perp})}{\delta j^e(\mathbf{z}_{\perp})} = \left[D_i\frac{1}{\partial D}\right]^{be}(\mathbf{x}_{\perp} ,\mathbf{z}_{\perp})\, $. 
Additionally using eq. \eqref{eq:Nperp_exp}, we calculate
\begin{equation}\label{eq:BL_BR}
Q_i^a({\mathbf x})\equiv [UN_{\perp} (b_L\mathcal{R}_p -b_R)]^a_i({\mathbf x})=-i\left[U({\mathbf x})\left(D_i\frac{1}{D^2} - \partial_i\frac{1}{\partial^2}\right)D\partial\right]^{ab}(\mathbf{x},\mathbf{z}) \frac{\delta }{\delta j^b(\mathbf{z})}\, .
\end{equation}
where the standard eikonal scattering matrix $U$ is defined in Eq.\eqref{u}.

Eq.\eqref{eq:phase_space_lindblad} now becomes
\begin{equation}\label{halfj}
\frac{d W[{\mathbf j}]}{dy}  = -\frac{1}{2\pi} \int_{\mathbf x}Q_i^{a\dagger}({\mathbf x)}Q_i^a{(\mathbf x})W[{\mathbf j}] 
\end{equation}

One can express the JIMWLK equation in terms of single gluon scattering matrix $U(\mathbf{x}_{\perp})^{ab}$ rather than the color charge density $j^a(\mathbf{x}_{\perp})$. The relation between the two was derived in \cite{Kovner:2007zu}.  Now we can relate functional derivatives with respect to the $U$ matrix to those with respect to the color current.  
\begin{equation}
\frac{\delta}{\delta j^a(\mathbf{x}_{\perp})} = \int d\mathbf{z}_{\perp} \frac{\delta U^{cd}(\mathbf{z}_{\perp})}{\delta j^a(\mathbf{x}_{\perp})} \frac{\delta}{\delta U^{cd}(\mathbf{z}_{\perp})}
\end{equation}
with
\begin{equation}
\begin{split}
 \frac{\delta U^{cd}(\mathbf{z}_{\perp})}{\delta j^a(\mathbf{x}_{\perp})} &=\int d^2\mathbf{y}_{\perp}  \frac{\delta U^{cd}(\mathbf{z}_{\perp})}{\delta b_l^b(\mathbf{y}_{\perp})}\frac{\delta b_l^b(\mathbf{y}_{\perp})}{\delta j^a(\mathbf{x}_{\perp})} \\
 & =ig \int_{\mathcal{C}} d^2 y_l \left(U^{\dagger bm}(\mathbf{y}_{\perp})\, T^{m}_{cn}\right) U^{nd}(\mathbf{z}_{\perp}) \left[D_l\frac{1}{\partial D}\right]^{ba}(\mathbf{y}_{\perp} , \mathbf{x}_{\perp})\\
 & = ig \int_{\mathcal{C}} d^2 y_l \left(U^{\dagger bm}(\mathbf{y}_{\perp})\, T^{m}_{cn}\right) U^{nd}(\mathbf{z}_{\perp}) \left[U^{\dagger} \partial_l U\frac{1}{\partial D}\right]^{ba}(\mathbf{y}_{\perp}  , \mathbf{x}_{\perp})\\
 &= igT^{e}_{cn}U^{nd}(\mathbf{z}_{\perp})  \int_{\mathcal{C}} d^2 y_l  \left[ \partial_l U\frac{1}{\partial D}\right]^{ea}(\mathbf{y}_{\perp}  , \mathbf{x}_{\perp})\\
 &= igT^{e}_{cn}U^{nd}(\mathbf{z}_{\perp})  U^{eb}(\mathbf{z}_{\perp})\left[ \frac{1}{\partial D}\right]^{ba}(\mathbf{z}_{\perp} ,\mathbf{x}_{\perp}) \\
 & = ig U^{cm}(\mathbf{z}_{\perp}) T^{m}_{db} \left[ \frac{1}{\partial D}\right]^{ba}(\mathbf{z}_{\perp}, \mathbf{x}_{\perp}) \\
 & =- ig\left[ U(\mathbf{z}_{\perp}) T^d \frac{1}{\partial D} \right]^{ca}(\mathbf{z}_{\perp} , \mathbf{x}_{\perp})
 \end{split}
\end{equation}
In the above we have used  $U^{\dagger} T^a U = U^{ab}T^b$ and $D_l = U^{\dagger}\partial_l U$. 
We have also used 
\begin{equation}
\begin{split}
\frac{\delta U^{cd}(\mathbf{z}_{\perp})}{\delta b_l^b(\mathbf{y}_{\perp})} =& U^{ce}(\mathbf{y}_{\perp})( ig T^b_{ef} )\int_{\mathcal{C}} d\mathbf{w}_l \delta(\mathbf{y}_{\perp} - \mathbf{w}_{\perp})[U^{\dagger}(\mathbf{y}_{\perp}) U(\mathbf{z}_{\perp})]^{fd}\\
=&ig \left(U^{\dagger bm}(\mathbf{y}_{\perp})\, T^{m}_{cn}\right) U^{nd}(\mathbf{z}_{\perp})\int_{\mathcal{C}} d\mathbf{w}_l \delta(\mathbf{y}_{\perp} - \mathbf{w}_{\perp})
\end{split}
\end{equation}
Finally, we obtain
\begin{equation}
\begin{split}
\frac{\delta}{\delta j^a(\mathbf{x}_{\perp})} &= \int d\mathbf{z}_{\perp} \frac{\delta U^{cd}(\mathbf{z}_{\perp})}{\delta j^a(\mathbf{x}_{\perp})} \frac{\delta}{\delta U^{cd}(\mathbf{z}_{\perp})}\\
&=ig\int d\mathbf{z}_{\perp}U^{cm}(\mathbf{z}_{\perp}) T^{m}_{db} \left[ \frac{1}{\partial D}\right]^{ba}(\mathbf{z}_{\perp}, \mathbf{x}_{\perp})\frac{\delta}{\delta U^{cd}(\mathbf{z}_{\perp})}\\
&=ig\int d\mathbf{z}_{\perp} \mathrm{Tr}\left(U(\mathbf{z}_{\perp}) T^b \frac{\delta}{\delta U^{\dagger}(\mathbf{z}_{\perp})}\right) \left[\frac{1}{\partial D} \right]^{ba}(\mathbf{z}_{\perp} , \mathbf{x}_{\perp})\\
&= -ig \int d\mathbf{z}_{\perp}  \left[\frac{1}{ D\partial} \right]^{ae}(\mathbf{x}_{\perp} , \mathbf{z}_{\perp})\mathcal{J}^e_R(\mathbf{z}_{\perp})\, .
\end{split}
\end{equation}
We have denoted
\begin{equation}
\mathcal{J}^b_R(\mathbf{z}_{\perp})  = -\mathrm{Tr}\left(U(\mathbf{z}_{\perp}) T^b \frac{\delta}{\delta U^{\dagger}(\mathbf{z}_{\perp})}\right)\, .
\end{equation}
Also recall that
\begin{equation}
\partial_i \frac{1}{\partial^2}(\mathbf{x}_{\perp}, \mathbf{y}_{\perp}) = \frac{1}{2\pi} \frac{(\mathbf{x}_{\perp}-\mathbf{y}_{\perp})_i}{(\mathbf{x}_{\perp}-\mathbf{y}_{\perp})^2}\,, \quad D_i \frac{1}{D^2}(\mathbf{x}_{\perp}, \mathbf{y}_{\perp}) = \frac{1}{2\pi}U^{\dagger}(\mathbf{x}_{\perp}) \frac{(\mathbf{x}_{\perp}-\mathbf{y}_{\perp})_i}{(\mathbf{x}_{\perp}-\mathbf{y}_{\perp})^2}U(\mathbf{y}_{\perp})\,. 
\end{equation}
Finally, eq. \eqref{eq:BL_BR} becomes
\begin{equation}\label{qu}
Q_i^a[\mathbf{x}_{\perp};U]  = -\frac{g}{2\pi}\int d\mathbf{y}_{\perp} \frac{(\mathbf{x}_{\perp}-\mathbf{y}_{\perp})_i}{(\mathbf{x}_{\perp}-\mathbf{y}_{\perp})^2} \left[ U(\mathbf{y}_{\perp}) -U(\mathbf{x}_{\perp})\right]^{ab} \mathcal{J}_R^b(\mathbf{y}_{\perp})
\end{equation}
The standard JIMWLK kernel is reproduced as
\begin{equation}
\begin{split}
& -\frac{1}{2\pi} \int d^2\mathbf{z}_{\perp} Q_i^{a\dagger}[\mathbf{z}_{\perp}; U] Q_i^a[\mathbf{z}_{\perp}; U] \\
=& - \frac{\alpha_s}{2\pi^2} \int_{\mathbf{z}_{\perp}, \mathbf{y}_{\perp},\mathbf{x}_{\perp}}\frac{(\mathbf{z}_{\perp} - \mathbf{x}_{\perp})_i}{(\mathbf{z}_{\perp} - \mathbf{x}_{\perp})^2}\frac{(\mathbf{z}_{\perp} - \mathbf{y}_{\perp})_i}{(\mathbf{z}_{\perp} - \mathbf{y}_{\perp})^2}\mathcal{J}^e_{R}(\mathbf{y}_{\perp})\\
&\times [ 1+ U^{\dagger}(\mathbf{y}_{\perp}) U(\mathbf{x}_{\perp}) -  U^{\dagger}(\mathbf{y}_{\perp}) U(\mathbf{z}_{\perp}) - U^{\dagger}(\mathbf{z}_{\perp}) U(\mathbf{x}_{\perp})]^{ed} \mathcal{J}^d_R(\mathbf{x}_{\perp})\, .
\end{split}
\end{equation}

\subsection{From JIMWLK back to Lindblad.}
Finally using Eq.\eqref{halfj} and Eq.\eqref{eq:BL_BR} we can transform the evolution equation back to Hilbert space. First we note, that the amplitude $Q_i^a$ is hermitian as an operator on the phase space. This is obvious from Eq.\eqref{qu}, since $\mathcal{J}_R^b(\mathbf{y}_{\perp})$ can be commuted to the left through the factors of $U$, as the right color index on $U$ is contracted with the index of $ \mathcal{J}_R^b(\mathbf{y}_{\perp})$. One can then write the JIMWLK equation as 
\begin{equation}
\frac{d W[{\mathbf j}]}{dy}  = -\frac{1}{2\pi} \int_{\mathbf x}\big[Q_i^{a}({\mathbf x)},\big[Q_i^a{(\mathbf x}),W[{\mathbf j}]\big] \big]
\end{equation}
where the commutator is understood as the commutator of the operators on phase space. Transforming this back to Hilbert space we see that to this order all we need to do is substitute $-i\frac{\delta }{\delta j^a(\mathbf{z})}\rightarrow \hat\Phi^a(\mathbf{z})$ and keep the structure of the double commutator. This procedure gives Eq.\eqref{qlin} as claimed.


\section{Discussion}

This paper is devoted to analysis of  the high energy limit of hadronic scattering, and its energy evolution formulated as effective quantum theory. 

The dynamics of this effective theory  is governed by a density matrix. Here we were able to define this "reduced" density matrix 
in a way reminiscent to an open quantum system, i.e. bi partitioning the degrees of freedom into the "system" and "environment" and integrating over the environment. In the present case the bi partitioning is into the ``valence" gluons as the system and ``soft" gluons as the environment. Despite some similarities,  there is a significant difference between the high energy limit considered here and bi partitioning in a quantum open system. In a quantum system one normally integrates completely the environment and then considers only observables that depend on the degrees of freedom of the "system". This is not  the case for the high energy limit, since the soft gluons contribute nontrivially to the color charge density, which is the basic observable in the effective theory. Defining the reduced density matrix is therefore rather nontrivial. Nevertheless we were able to do it.


We have then followed the usual assumption made in the derivation of the high energy evolution, i.e. that only the distribution of the color charge density in the transverse plane is relevant for determining hadronic properties at high energy. Assuming that the reduced density matrix of a hadronic system depends only on the components of the color charge density
results in a quasi diagonal density matrix in the sense that its matrix elements between states belonging to different color representation vanish. This is true both in dense and dilute limits, and generalizes the notion of diagonal density matrix discussed in \cite{Armesto:2019mna} to arbitrary parametric values of color charge density.

Under this assumption we have shown that the rapidity evolution of the reduced density matrix is of the Lindblad type in the two limiting cases - the dense (JIMWLK) and the dilute (KLWMIJ) limits. This is true even though the nature of the energy evolution is in principle quite different from the nature of time evolution of a dynamical quantum system. 

Interestingly the evolution equation that interpolates between the two limits, Eq.\eqref{eq:nonlindblad_eq_cgc} does not have a Lindblad form. Although the derivation of this interpolating equation is not under parametric control, the basic features of the derivation are generic, and our analysis shows that the absence of Lindblad form should be a rule rather than exception. The basic reason is that the rapidity plays a dual role in high energy evolution: it is the analog of the evolution time on one hand, and  is a quantum number that labels the quantum states of the environment that are integrated out on the other hand. This invalidates in principle the usual argument for the Lindblad form of the differential evolution equation.

Finally, we have shown how to rigorously relate the reduced density matrix description of the evolution with the approach used in most pertinent literature based on the probability density functional $W[\mathbf{j}]$.
To this end we have  explored the Wigner-Weyl transformation, which maps the Hilbert space description of a quantum system in terms of density matrix $\hat\rho$ into the classical phase space description in terms of quasi probability distribution $W$. By adapting this transformation to the present case we have shown that the Lindblad evolution equation for $\hat\rho$ is indeed equivalent to a Fokker-Planck type equation for $W$, where the components of the color charge density $\mathbf{j}$ are considered as coordinates on a classical non Abelian phase space. This Fokker-Planck equation reduces to KLWMIJ and JIMWLK equations in the appropriate dilute and dense limits.

In quantum optics, it has been known for decades that the Lindblad master equation maps to the Fokker-Planck equation through quantum-classical correspondence. Here we have established the same in the context of high energy evolution with the JIMWLK (or KLWMIJ) playing the role of the Focker-Planck equation.

We stress again that this paper deals only with the conventional JIMWLK/KLWMIJ setup, where the density matrix is assumed to depend only on the color charge density degrees of freedom. Recently it was suggested that this framework may be too restrictive and may not be adequate for studying some interesting observables at high energy \cite{Armesto:2019mna}. Such observables, like correlations between the transverse momentum and density in the transverse plane may be formally subleading at high energy, but could be of great interest in the study of correlations in particle production.  In order to include these into consideration one has to extend the conventional framework and allow for density matrices that depend not just on color charge density $\hat{\mathbf{j}}$, but also on their conjugate variables, which in the present paper we have identified as the operators $\hat\Phi^a$. It was suggested in \cite{Armesto:2019mna} that the evolution of this more general density matrix is also given by the same Lindblad equation. Although this seems very likely to be the case, our current derivation does not cover this interesting more general situation. It should be possible to extend our current method to deal with this intriguing problem. This investigation is currently under way.

We now comment on several questions/issues that arise from our results. 

First, the fact that the evolution of the reduced density matrix beyond dense-dilute limit is most likely not of Lindblad type
 begs an interesting general question. It is known that Lindblad equation preserves the properties of the density matrix, namely normalization and positivity. Is this also the case for Eq.\eqref{eq:nonlindblad_eq_cgc} even though it is not in Lindblad form? It is quite obvious that the normalization of the density matrix is preserved under Eq.\eqref{eq:nonlindblad_eq_cgc}, since its right hand side is a commutator, and therefore has a vanishing trace. As for the positivity, it is  more difficult to establish. We note however, that the differential evolution follows from the Krauss representation Eq.\eqref{eq:cgc_kraus_rep} which does preserve positivity \cite{Preskill:2019}. We therefore believe that the differential evolution Eq.\eqref{eq:nonlindblad_eq_cgc} does indeed preserve positivity and thus is  a consistent evolution of a density matrix. If this is the case, one is lead to a general conclusion that the set of possible differential evolutions of a density matrix is not limited to equations of Lindblad type.

Second, we note that one of the useful perspectives on the JIMWLK evolution is that of  a Langevin equation for Brownian motion in the space of Wilson line $U^{ab}(\mathbf{x}_{\perp})$ \cite{Weigert:2000gi, Blaizot:2002np}. The bi partitioning into the ``system" of the hard gluons and ``environment" due to the soft gluons harmonizes nicely with the random walk picture. After a boost by $\Delta y$, the soft gluons can be emitted into any of the multigluon  Fock states $\{ |n\rangle \}$. This emission contributes to  a random addition to the color charge density  $j^a_{y+\Delta y} \sim j^a_{y} + \delta j^a$ with $\delta j^a$ being a random variable, which therefore random walks in the color space.   
The Langevin equation is a reformulation of the Focker-Planck equation, which is equivalent to JIMWLK. It is then interesting to ask whether such a Langevin description  can be extended beyond the leading order. The NLO JIMWLK equation has been derived some years ago \cite{nlo:balitsky, nlo:kovner, nlo:lublinsky, nlo:caron-huot}. Naturally the derivation involves integration over gluon and quark degrees of freedom in the rapidity interval $\Delta y$. As opposed to the leading order, where as we discussed the probability to create all single gluon states is equal, independent of their rapidities, at NLO there is a genuine integration over rapidity of two soft parton states. This suggests that the evolution equation for the density matrix is not of Lindblad type for the same reason Eq.\eqref{eq:nonlindblad_eq_cgc} is not.  If that is the case one does not expect it to be equivalent to a Focker-Planck equation for the quasi probability function and thus the Langevin description may well not be possible.

We hope that the new perspective on high energy evolution discussed in this paper will be useful not only for a more fundamental understanding of JIMWLK equation but will also prove useful for future developments.

\appendix

\section{\boldmath The Operator $\hat{\Phi}^a$}
In this Appendix we present the derivation of the quantum ``phase" operator $\hat\Phi^a$, which via eq.\eqref{R} defines the quantum shift operator of the color charge density. Part of this derivation appears in the text, but we keep here all the details for completeness.

We are looking for $M^{ab}(\Phi)$, as a functinal of operator $\hat{\Phi}^a$ that satisfies the commutation relations 
\begin{equation}
\begin{split}
&[\hat{\Phi}^a, \hat{\Phi}^b] =0\, ,\\
&[\hat{\Phi}^a, \hat{j}^b] = M^{ab}(\Phi)\, \\
\end{split}
\end{equation}
so that the color charge density shift operation is  
\begin{equation}\label{eq:shift_requirement1}
\mathrm{exp}\left\{i\hat{j}^a_{\mathrm{soft}} \hat{\Phi}^a\right\} \hat{j}^e\, \mathrm{exp}\left\{-i\hat{j}^a_{\mathrm{soft}} \hat{\Phi}^a\right\} = \hat{j}^e + \hat{j}^e_{\mathrm{soft}}\, .
\end{equation}
Using the Baker-Hausdaurff formula
\begin{equation}
e^X Y e^{-X} = Y + [X, Y] + \frac{1}{2!}[X, [X, Y]] + \frac{1}{3!}[X,[X,[X,Y]]]+\ldots + \frac{1}{n!}[X, [X, [\ldots [X, Y]\ldots]]] + \ldots
\end{equation}
and expanding eq.\eqref{eq:shift_requirement1} in commutators,  the first three terms are
\begin{equation}
[i\hat{j}^a_{\mathrm{soft}} \hat{\Phi}^a, \hat{j}^e]   = i\hat{j}^a_{\mathrm{soft}} M^{ae}(\Phi)\, ,
\end{equation}
\begin{equation}
\begin{split}
\frac{1}{2!}[i\hat{j}^b_{\mathrm{soft}} \hat{\Phi}^b, [i\hat{j}^a_{\mathrm{soft}} \hat{\Phi}^a, \hat{j}^e]   ] &= \frac{1}{2!} [\hat{j}^b_{\mathrm{soft}}, \hat{j}^a_{\mathrm{soft}}] i^2 \hat{\Phi}^b M^{ae}(\Phi) = \frac{1}{2!} igf^{bac} \hat{j}^{c}_{\mathrm{soft}} i^2 \hat{\Phi}^b M^{ae}(\Phi) \\
&= \frac{1}{2!}\,  i\hat{j}^c_{\mathrm{soft}} \left(igT^b_{ca} \hat{\Phi}^b \right)M^{ae}(\Phi)\\
&= \frac{1}{2!} \, i\hat{j}^a_{\mathrm{soft}} \left(igT^b \hat{\Phi}^b M(\Phi)\right)_{ae}\, ,\\
\end{split}
\end{equation}
\begin{equation}
\frac{1}{3!}[i\hat{j}^c_{\mathrm{soft}} \hat{\Phi}^c,[i\hat{j}^b_{\mathrm{soft}} \hat{\Phi}^b, [i\hat{j}^a_{\mathrm{soft}} \hat{\Phi}^a, \hat{j}^e]   ] ] = \frac{1}{3!} i\hat{j}^a_{\mathrm{soft}} \left((igT^b \hat{\Phi}^b)^2 M(\Phi)\right)_{ae}\, .
\end{equation}
Here we have used $-if^{abc} = T^a_{bc}$.  

From the above explicit calculations, it is natural to make the following ansatz
\begin{equation}
M^{ab}(\Phi) = -i\sum_{n=0}^{\infty} c_n \left[\chi^n\right]_{ab}\, , \quad\mathrm{with}\,\,\,  \chi = igT^b\hat{\Phi}^b\, .
\end{equation}
Clearly, $c_0 =1$ follows from the requirement eq.\eqref{eq:shift_requirement1}. This requirement further imposes the constraint
\begin{equation}
i + \sum_{k=0}^{\infty} \frac{1}{(k+1)!} \left[\chi^k M(\chi)\right]^{ab}=0
\end{equation}
which after substituting the ansatz for $M(\chi)$ becomes
\begin{equation}
1= \sum_{k=0}^{\infty}\sum_{m=0}^{\infty}\frac{c_m}{(k+1)!} \chi^{k+m}_{ab}\, .
\end{equation}
Note that for $k=0, m=0$, the $c_0=1$ automatically satisfies the above condition. For $N=k+m\geq 1$,  the coefficients of $\chi^N$ have to be vanishing, one then obtains
\begin{equation}
\sum_{m=0}^N \frac{c_m}{(N-m+1)!} =0
\end{equation}
which is equivalent to the following recursive relations 
\begin{equation}\label{eq:recursive_relations1}
c_N = -\sum_{m=0}^{N-1} \frac{c_m}{(N-m+1)!} \, , \quad \mathrm{with}\,\, c_0=1.
\end{equation}
A few examples can be explicitly calculated
\begin{equation}
\begin{split}
& c_1 = -\frac{c_0}{2!} = -\frac{1}{2} \,, \\
&c_2 =  - \frac{c_0}{3!} - \frac{c_1}{2!} =  \frac{1}{12} \, , \\
&c_3 = -\frac{c_0}{4!}  - \frac{c_1}{3!} - \frac{c_2}{2!} = 0\, ,\\
&c_4 = -\frac{c_0}{5!} - \frac{c_1}{4!} -\frac{c_2}{3!} -\frac{c_3}{2!} = - \frac{1}{6!} = -\frac{1}{720}\,, \\
&c_5 = -\frac{c_0}{6!} -\frac{c_1}{5!} -\frac{c_2}{4!}-\frac{c_4}{2!} =0\, , \\
&c_6 = -\frac{c_0}{7!} -\frac{c_1}{6!} -\frac{c_2}{5!} -\frac{c_4}{3!} = \frac{1}{6}\frac{1}{7!} =\frac{1}{30240}\, ,\\
&c_7 = -\frac{c_0}{8!} -\frac{c_1}{7!} -\frac{c_2}{6!} -\frac{c_4}{4!}-\frac{c_6}{2!} = 0 \, , \\
&c_8 = -\frac{c_0}{9!} -\frac{c_1}{8!} -\frac{c_2}{7!} -\frac{c_4}{5!}-\frac{c_6}{3!} =  - \frac{1}{30}\frac{1}{8!} = -\frac{1}{1209600}\, .
\end{split}
\end{equation}
It turns out that these numbers correspond to the coefficients in expanding the function
\begin{equation}\label{eq:M_expression1}
M^{ab}(\chi) = -i \left[\frac{\chi}{2} \coth{\frac{\chi}{2}}  - \frac{\chi}{2}\right]^{ab}\, .
\end{equation}
One can now explicitly check that  Taylor expansion of eq. \eqref{eq:M_expression1} in $\chi$ reproduces all the coefficients calculated using the recursive relations in eq. \eqref{eq:recursive_relations1}.  

In addition one needs to check the consistence of the Lie algebra constructed from $\hat{\Phi}^a$ and $\hat{j}^a$. This consistency requires that the following Jacobi identity holds
\begin{equation}
[[\hat{\Phi}^a, \hat{j}^b], \hat{j}^c] + [[\hat{j}^b, \hat{j}^c],\hat{\Phi}^a] + [[\hat{j}^c, \hat{\Phi}^a], \hat{j}^b] =0\, ,
\end{equation}
which is equivalent to
\begin{equation}\label{eq:consistence_check1}
[M^{ab}, \hat{j}^c] - [M^{ac}, \hat{j}^b] = igf^{bcd} M^{ad}\, .
\end{equation}
On the left hand side
\begin{equation}
\begin{split}
[M^{ab}, \hat{j}^c] &= -i\sum_{n=0}^{\infty}c_n [\chi^n_{ab}, \hat{j}^c]\\
&=-i\sum_{n=1}^{\infty} c_n \sum_{k=0}^{n-1} \chi^k_{ae_1}\, [\chi_{e_1e_2}, \hat{j}^c] \, \chi^{n-1-k}_{e_2b}\\
&=-i\sum_{n=1}^{\infty} c_n \sum_{k=0}^{n-1} \chi^k_{ae_1}\, ig T^d_{e_1e_2}[\hat{\Phi}^d, \hat{j}^c] \, \chi^{n-1-k}_{e_2b}\\
&=-i\sum_{n=1}^{\infty} c_n \sum_{k=0}^{n-1} \chi^k_{ae_1}\, ig T^d_{e_1e_2} \, \chi^{n-1-k}_{e_2b} M^{dc}\\
&=(-i)^2\sum_{n=1}^{\infty} c_n \sum_{k=0}^{n-1} \chi^k_{ae_1}\, ig T^d_{e_1e_2} \, \chi^{n-1-k}_{e_2b} \sum_{m=0}^{\infty} c_m \chi^m_{dc}\, .\\
\end{split}
\end{equation}
On the right hand side
\begin{equation}
igf^{bcd}M^{ad} = \sum_{N=0}^{\infty} gf^{bcd}c_N \chi^N_{ad}\, .
\end{equation}
We here check the consistency condition eq. \eqref{eq:consistence_check1} order by order to verify that it indeed is satisfied. We do not have a proof for the case of general N, but we believe the same procedure can be carried out to high order terms.

 The left hand side of eq.\eqref{eq:consistence_check1} has $N= m+n-1$ in terms of power of $\hat{\Phi}^a$. Also note that $m\geq 0$ and $n\geq 1$.   In the following, we calculate the cases $N=0, 1, 2, 3$ in details.

\subsection*{\boldmath $N=0$}
For $N=0$, the right hand side is $gf^{bcd} c_0\delta_{ad} = gf^{bca}$, the left hand side can have $n=1, m=0$
\begin{equation}
\begin{split}
&(-i)^2 c_0c_1 \delta_{ae_1}\left( igT^{d}_{e_1e_2}\right)\delta_{e_2b} \delta_{dc} -(b\leftrightarrow c)\\
=&\frac{1}{2} igT^c_{ab} -(b\leftrightarrow c) = \frac{1}{2} ig (-if^{cab}) - (b\leftrightarrow c)\\
=& gf^{abc}\, .
\end{split}
\end{equation}
For $N=0$ terms eq.\eqref{eq:consistence_check1} holds.

\subsection*{\boldmath $N=1$}
For $N=1$,  the right hand side becomes
\begin{equation}
gf^{bcd} c_1 \chi_{ad} = -\frac{1}{2} gf^{bcd} (igT^e_{ad} \Phi^e) = \frac{1}{2}g^2 T^d_{bc} T^e_{ad} \Phi^e\, .
\end{equation}
For the left hand side $N=m+n-1 =1$, we have two possibilities $n=1, m=1$ and $n=2, m=0$. 
\begin{equation}
\begin{split}
&- c_0c_2 \sum_{k=0}^1 \chi_{ae_1}^k (igT^d_{e_1e_2}) \chi^{1-k}_{e_2b} \delta_{dc} - c_1c_1 \delta_{ae_1} (igT^d_{e_1e_2} )\delta_{e_2b} \chi_{dc} - (b\leftrightarrow c)\\
=&-c_0c_2[\delta_{ae_1} (igT^d_{e_1e_2}) (igT^e_{e_2b}\Phi^e) \delta_{dc} + (igT^e_{ae_1}\Phi^e) (igT^d_{e_1e_2}) \delta_{e_2b}\delta_{dc}] -c_1^2(igT^d_{ab})(igT^e_{dc}\Phi^e) - (b\leftrightarrow c)\\
=&-c_0c_2 (ig)^2 [-T^c_{ae_2} T^b_{e_2e}+T^e_{ae_1}T^c_{e_1b}]\Phi^e -c_1^2(ig)^2 T^b_{ad}T^c_{de}\Phi^e - (b\leftrightarrow c)\\
=&-c_0c_2(ig)^2[-(T^cT^b)_{ae} + T^e_{ae_1}T^{e_1}_{bc}]\Phi^e - c_1^2 (ig)^2 (T^bT^c)_{ae} \Phi^e -(b\leftrightarrow c)\\
=&-c_0c_2(ig)^2[-if^{cbd}T^d_{ae} + 2 T^e_{ae_1}T^{e_1}_{bc}]\Phi^e - c_1^2 (ig)^2 if^{bcd}T^d_{ae}\Phi^e\\
=&-g^2\left(-3c_0c_2-c_1^2\right)T^d_{bc}T^e_{ad}\Phi^e = \frac{1}{2} g^2 T^d_{bc}T^e_{ad}\Phi^e\, .\\
\end{split}
\end{equation}
Therefore for $N=1$ terms eq.\eqref{eq:consistence_check1} holds.

\subsection*{\boldmath $N=2$}
For $N=2$, the right hand side becomes
\begin{equation}
gf^{bcd} c_2 \chi^2_{ad} = (ig)^3 c_2 T^d_{bc}(T^{e_1}T^{e_2})_{ad}\Phi^{e_1}\Phi^{e_2}\, .
\end{equation}
For the left hand side $N=m+n-1=2$, there are three possibilities $n=1, m=2$; $n=2, m=1$; $n=3, m=0$.  From $c_3=0$, we can only consider the first two possibilities. 
\begin{equation}
\begin{split}
&(-i)^2 c_1c_2 \sum_{k=0}^1 \chi^k_{ae_1} (igT^d_{e_1e_2})\chi^{1-k}_{e_2b} (igT^{e_3}_{dc}\Phi^{e_3}) + (-i)^2 c_2c_1 \delta_{ae_1}(igT^d_{e_1e_2})(igT^{e_3}\Phi^{e_3} igT^{e_4}\Phi^{e_4})_{dc}\delta_{e_2b}\\
=&-(ig)^3 c_1c_2 (T^d_{ae_2}T^{e_3}_{dc}T^{e_4}_{e_2b} + T^{e_4}_{ae_1} T^d_{e_1b}T^{e_3}_{dc})\Phi^{e_3}\Phi^{e_4} - (ig)^3 c_2c_1 T^d_{ab}(T^{e_3}T^{e_4})_{dc} \Phi^{e_3}\Phi^{e_4}\\
=&-(ig)^3 c_1c_2[ -(T^bT^aT^c)_{e_4e_3} + (T^cT^bT^a)_{e_3e_4} + (T^bT^{e_3}T^c)_{ae_4}] \Phi^{e_3}\Phi^{e_4}\\
=&-(ig)^3 c_1c_2[ 2 (T^cT^bT^a)_{e_3e_4} -(T^bT^{e_4}T^c)_{ae_3}+ (T^bT^{e_3}T^c)_{ae_4}] \Phi^{e_3}\Phi^{e_4}
\end{split}
\end{equation}
Note that $e_3$ and $e_4$ are symmetric. The last two terms in last equality cancel.  After subtracting the $(b\leftrightarrow c)$ part, one obtains
\begin{equation}
-(ig)^3 c_1c_2 2 if^{cbe}(T^eT^a)_{e_3e_4}\Phi^{e_3}\Phi^{e_4} =- (ig)^3 2c_1c_2 T^e_{bc}(T^{e_3}T^{e_4})_{ae} \Phi^{e_3}\Phi^{e_4} 
\end{equation} 
Since  $c_1 = -1/2$, clearly for $N=2$ terms eq.\eqref{eq:consistence_check1} holds.

\subsection*{\boldmath $N=3$}
For $N=3$ the right hand side vanishes because $c_3 =0$.  We have to show that the left hand side also vanishes. First note that for $N=m+n-1=3$, there are four possibilities: $(n=1, m=3)$; $(n=2, m=2)$; $(n=3, m=1)$ ; $(n=4, m=0)$. Only the two cases $n=2,m=2$ and $n=4, m=0$ contribute. We then need to show that the following terms cancel.
\begin{equation}
\begin{split}
&-c_2^2 \Big[(T^dT^{d_3})_{ab}(T^{d_1}T^{d_2})_{dc} + (T^{d_3}T^d)_{ab}(T^{d_1}T^{d_2})_{dc}\Big] -c_0c_4 \Big[(T^cT^{d_1}T^{d_2}T^{d_3})_{ab} \\
&+ (T^{d_1}T^cT^{d_2}T^{d_3})_{ab}+(T^{d_1}T^{d_2}T^cT^{d_3})_{ab} +(T^{d_1}T^{d_2}T^{d_3}T^c)_{ab}\Big] -(b\leftrightarrow c)\\
\end{split}
\end{equation}
The guiding principles of organizing these terms are the following:
\begin{itemize}
\item $d_1, d_2, d_3$ are symmetric indices, we are free to interchange among them.
\item We rearrange terms according to their $b, c$ indices. If $b,c$ are indices of the same  matrix like $T^m_{bc}$, there  is no need for further simplication. If we have $T^b, T^c$ in adjacent  position like $T^bT^c$,  the $b\leftrightarrow c$ subtraction gives a commutator, which results in placing $b,c$ indices on the same matrix $T$. Then no further simplification is needed. 
\item If $T^b, T^c$ are not directly in the adjacent position, we rearrange the $b,c$ as the indices of matrix element not the label of $T$ matrix. 
\item We use the cancellations between terms like $(T^aT^{d_1}T^{d_2}T^{d_3})_{bc}$ and $(T^{d_1}T^{d_2}T^{d_3}T^a)_{cb}$.
\end{itemize}
First note the last two terms in the second bracket
\begin{equation}
(T^{d_1}T^{d_2}T^{d_3}T^c)_{ab} -(b\leftrightarrow c) = 2 (T^{d_1}T^{d_2}T^{d_3})_{ae}T^{e}_{bc}\, 
\end{equation}
and 
\begin{equation}
\begin{split}
&(T^{d_1}T^{d_2}T^cT^{d_3})_{ab} -(b\leftrightarrow c)\\
=&-(T^{d_1}T^{d_2}T^cT^b)_{ad_3} - (b\leftrightarrow c)\\
=&(T^{d_1}T^{d_2}T^{d_3})_{ae}T^e_{bc}\, .\\
\end{split}
\end{equation}
the first two terms in the second bracket
\begin{equation}
\begin{split}
&(T^cT^{d_1}T^{d_2}T^{d_3})_{ab} + (T^{d_1}T^cT^{d_2}T^{d_3})_{ab} - (b\leftrightarrow c)\\
=&2(T^cT^{d_1}T^{d_2}T^{d_3})_{ab} -T^e_{d_1c}(T^eT^{d_2}T^{d_3})_{ab} - (b\leftrightarrow c)\\
=&-2(T^aT^{d_1}T^{d_2}T^{d_3})_{cb} + (T^{d_1}T^aT^{d_2}T^{d_3})_{cb} +2(T^aT^{d_1}T^{d_2}T^{d_3})_{bc} - (T^{d_1}T^aT^{d_2}T^{d_3})_{bc} \\
=&-2if^{ad_1e}(T^eT^{d_2}T^{d_3})_{cb} - 2if^{ad_2e}(T^{d_1}T^eT^{d_3})_{cb} - 2if^{ad_3e}(T^{d_1}T^{d_2}T^e)_{cb} + if^{ad_2e}(T^{d_1}T^eT^{d_3})_{cb}\\
=&- 5if^{ad_2e}(T^{d_1}T^eT^{d_3})_{cb}  - 2if^{ad_1e}if^{ed_2 h}(T^hT^{d_3})_{cb} -2if^{ad_3e}if^{d_2eh}(T^{d_1}T^h)_{cb}\\
=&- 5if^{ad_2e}(T^{d_1}T^eT^{d_3})_{cb} - 2if^{ad_1e}if^{ed_2h} if^{hd_3 d}T^d_{cb}\\
=&- 5if^{ad_2e}(T^{d_1}T^eT^{d_3})_{cb}+2(T^{d_1}T^{d_2}T^{d_3})_{ad}T^{d}_{bc}\\
\end{split}
\end{equation}
In obtaining the third equality, we have moved the $T^a$ matrix in the first term gradually to the far right so that the resulting term will cancel the third term. We also moved the $T^a$ matrix in the second term passing $T^{d_2}$, which then will cancel the fourth term. 
Now consider the terms in the first bracket.
\begin{equation}
\begin{split}
&(T^dT^{d_3})_{ab}(T^{d_1}T^{d_2})_{dc} + (T^{d_3}T^d)_{ab}(T^{d_1}T^{d_2})_{dc} - (b\leftrightarrow c)\\
=&-(T^{d_3}T^aT^{d_1}T^{d_2})_{bc} - (T^{d_3}T^bT^{d_1}T^{d_2})_{ac} - (b\leftrightarrow c)\\
=&-(T^{d_3}T^aT^{d_1}T^{d_2})_{bc} - (T^bT^{d_3}T^{d_1}T^{d_2})_{ac} +T^e_{d_3 b}(T^eT^{d_1}T^{d_2})_{ac} - (b\leftrightarrow c)\\
=&-2(T^{d_3}T^aT^{d_1}T^{d_2})_{bc} + (T^aT^{d_3}T^{d_1}T^{d_2})_{bc} +2(T^{d_3}T^aT^{d_1}T^{d_2})_{cb} - (T^aT^{d_3}T^{d_1}T^{d_2})_{cb} \\
=&-2if^{ad_1e}(T^{d_3}T^eT^{d_2})_{bc} + if^{ad_3e}(T^eT^{d_1}T^{d_2})_{bc} + if^{ad_1e}(T^{d_3}T^eT^{d_2})_{bc} + if^{ad_2e}(T^{d_3}T^{d_1}T^{e})_{bc}\\
=&if^{ad_1e}(T^{d_3}T^eT^{d_2}) _{bc}+ if^{ad_3e}if^{ed_1 h}(T^hT^{d_2})_{bc} + if^{ad_2e} if^{d_1eh}(T^{d_3}T^h)_{bc}\\
=&if^{ad_1e}(T^{d_3}T^eT^{d_2})_{bc} + if^{ad_3e}if^{ed_1h} if^{hd_2 d}T^d_{bc}\\
=&if^{ad_1e}(T^{d_3}T^eT^{d_2})_{bc} + (T^{d_3}T^{d_1}T^{d_2})_{ad}T^d_{bc}\\
\end{split}
\end{equation}
Note that $c_2^2 = (1/12)^2 = \frac{5}{6!}$ and $c_0c_4 = -\frac{1}{6!}$.  Therefore for $N=3$ terms eq.\eqref{eq:consistence_check1} holds.

We have checked four leading terms in the expansion of the Jacobi identity. The same explicit procedure can be followed for higher order terms as well, but it becomes increasingly cumbersome. We therefore stop at this point.

\acknowledgments

We thank Nestor Armesto and Michael Lublinsky for interesting and useful discussion on the subject of this paper.
The work is supported by the NSF Nuclear Theory grant 1913890.


\begin{thebibliography}{99}


 
  \bibitem{Balitsky:1995ub} 
  I.~Balitsky,
  \emph{``Operator expansion for high-energy scattering''},
  \emph{Nucl.\ Phys.\ B} {\bf 463}, 99 (1996)


  \bibitem{JalilianMarian:1997jx} 
  J.~Jalilian-Marian, A.~Kovner, A.~Leonidov and H.~Weigert,
 \emph{``The BFKL equation from the Wilson renormalization group''},
  \emph{Nucl.\ Phys.\ B }{\bf 504}, 415 (1997)
  
  \bibitem{JalilianMarian:1997gr} 
  J.~Jalilian-Marian, A.~Kovner, A.~Leonidov and H.~Weigert,
  \emph{``The Wilson renormalization group for low x physics: Towards the high density regime''},
  \emph{Phys.\ Rev.\ D} {\bf 59}, 014014 (1998)
  
  \bibitem{JalilianMarian:1997dw} 
  J.~Jalilian-Marian, A.~Kovner and H.~Weigert,
  \emph{``The Wilson renormalization group for low x physics: Gluon evolution at finite parton density''},
  \emph{Phys.\ Rev.\ D} {\bf 59}, 014015 (1998)
 
 
 
  \bibitem{Kovner:2000pt} 
  A.~Kovner, J.~G.~Milhano and H.~Weigert,
  \emph{``Relating different approaches to nonlinear QCD evolution at finite gluon density''},
  \emph{Phys.\ Rev.\ D} {\bf 62}, 114005 (2000)

   
  \bibitem{Iancu:2000hn} 
  E.~Iancu, A.~Leonidov and L.~D.~McLerran,
  \emph{``Nonlinear gluon evolution in the color glass condensate. 1''},
  \emph{Nucl.\ Phys.\ A} {\bf 692}, 583 (2001)
  
  \bibitem{Ferreiro:2001qy} 
  E.~Ferreiro, E.~Iancu, A.~Leonidov and L.~McLerran,
  \emph{``Nonlinear gluon evolution in the color glass condensate. 2''},
  \emph{Nucl.\ Phys.\ A} {\bf 703}, 489 (2002)
     
   \bibitem{Mueller:2001uk} 
  A.~H.~Mueller,
  \emph{``A Simple derivation of the JIMWLK equation''},
  \emph{Phys.\ Lett.\ B} {\bf 523}, 243 (2001)
  


\bibitem{Armesto:2019mna} N. Armesto, F. Dominguez, A. Kovner, M
                        Lublinsky and V. Skokov,  
  \emph{"The Color Glass Condensate density matrix: Lindblad evolution, entanglement entropy and Wigner functional"},
 \emph{JHEP}\ 05 (2019) 025;
                        
\bibitem{Kovner:2005uw} 
  A.~Kovner and M.~Lublinsky,
  \emph{``Dense-dilute duality at work: Dipoles of the target''}, 
  \emph{Phys.\ Rev.\ D} {\bf 72}, 074023 (2005)
  


 \bibitem{Gorini:1975nb} 
  V.~Gorini, A.~Kossakowski and E.~C.~G.~Sudarshan,
  \emph{``Completely Positive Dynamical Semigroups of N Level Systems''}, 
  \emph{J.\ Math.\ Phys.}\  {\bf 17}, 821 (1976).
  
 \bibitem{Lindblad:1975ef} 
  G.~Lindblad,
  \emph{``On the Generators of Quantum Dynamical Semigroups''},
 \emph{ Commun.\ Math.\ Phys.\ } {\bf 48}, 119 (1976).
 


   
  
 \bibitem{Altinoluk:2009je} 
  T.~Altinoluk, A.~Kovner, M.~Lublinsky and J.~Peressutti,
  \emph{``QCD Reggeon Field Theory for every day: Pomeron loops included''}, 
  \emph{JHEP} {\bf 0903}, 109 (2009)
 
 \bibitem{Hillery:1983ms} 
  M.~Hillery, R.~F.~O'Connell, M.~O.~Scully and E.~P.~Wigner,
  \emph{``Distribution functions in physics: Fundamentals''},
  \emph{Phys.\ Rept.}\  {\bf 106}, 121 (1984).
 
 \bibitem{Preskill:2019}
 J.~Preskill, 
 \emph{ ``Lecture notes for physics 219: Quantum
computation”},  available at $http://www.theory.caltech.edu/~preskill/ph219/ph219\_2018-19$, 2019,

 
 \bibitem{Carmichael:1993}
 H. ~J.~Carmichael, 
 \emph{``An Open Systems Approach to Quantum Optics ''}, 
 Springer (1993), Berlin.
 
 \bibitem{Breuer:2007}
 H. ~P.~Breuer,
 \emph{``The Theory of Open Quantum Systems ''} , 
 Oxford University Press, USA(March 29, 2007).
 
  \bibitem{Kovchegov:2012mbw} 
  Y.~V.~Kovchegov and E.~Levin,
  \emph{``Quantum chromodynamics at high energy''},
  Camb.\ Monogr.\ Part.\ Phys.\ Nucl.\ Phys.\ Cosmol.\  {\bf 33}, 1 (2012).
  

 

 \bibitem{McLerran:1993ni} 
  L.~D.~McLerran and R.~Venugopalan,
  \emph{``Computing quark and gluon distribution functions for very large nuclei''},
  \emph{Phys.\ Rev.\ D} {\bf 49}, 2233 (1994);
  
  \bibitem{McLerran:1993ka} 
  L.~D.~McLerran and R.~Venugopalan,
  \emph{``Gluon distribution functions for very large nuclei at small transverse momentum''},
  \emph{Phys.\ Rev.\ D} {\bf 49}, 3352 (1994)
  
  
  \bibitem{Kovner:2005pe} 
  A.~Kovner,
  \emph{``High energy evolution: The Wave function point of view''},
  \emph{Acta Phys.\ Polon.\ B} {\bf 36}, 3551 (2005)
  
 \bibitem{Kovner:2007zu} 
  A.~Kovner, M.~Lublinsky and U.~Wiedemann,
  \emph{``From bubbles to foam: Dilute to dense evolution of hadronic wave function at high energy''},
  \emph{JHEP} {\bf 0706}, 075 (2007)
  
 \bibitem{Kovner:2005nq} 
  A.~Kovner and M.~Lublinsky,
  \emph{``In pursuit of Pomeron loops: The JIMWLK equation and the Wess-Zumino term''}
  \emph{Phys.\ Rev.\ D} {\bf 71}, 085004 (2005)   
  
  
   
  \bibitem{Cahill:1969iq} 
  K.~E.~Cahill and R.~J.~Glauber,
  \emph{``Density operators and quasi probability distributions''}, 
  \emph{Phys.\ Rev.\ } {\bf 177}, 1882 (1969).
  

\bibitem{Agarwal:1971wc} 
  G.~S.~Agarwal and E.~Wolf,
  \emph{``Calculus for functions of noncommuting operators and general phase-space methods in quantum mechanics. i. mapping theorems and ordering of functions of noncommuting operators''}, 
  \emph{Phys.\ Rev.\ D} {\bf 2}, 2161 (1970).
  
 \bibitem{Agarwal:1971wb} 
  G.~S.~Agarwal and E.~Wolf,
  \emph{``Calculus for functions of noncommuting operators and general phase-space methods in quantum mechanics. ii. quantum mechanics in phase space''}, 
  \emph{Phys.\ Rev.\ D }{\bf 2}, 2187 (1970).
  
 
 \bibitem{Stratonovich:1956}
 R. ~L. ~Stratonovich, 
 \emph{Zh. Eksp. Teor, Fiz}. {\bf 31}, 1012 (1956),
 [\emph{Sov. Phys. JETP} {\bf 4}, 891 (1957)]. 
 
 \bibitem{Varilly:1989sv} 
  J.~C.~Varilly and J.~M.~Gracia-Bondia,
  \emph{``The Moyal representation for spin''},
  \emph{Annals Phys.\ } {\bf 190}, 107 (1989).
 
 \bibitem{Brif:1997km} 
  C.~Brif and A.~Mann,
  \emph{``A general theory of phase space quasi probability distributions''},
  \emph{J.\ Phys.\ A} {\bf 31}, L9 (1998)
 
 
 \bibitem{Brif:1998pw} 
  C.~Brif and A.~Mann,
  \emph{``Phase space formulation of quantum mechanics and quantum state reconstruction for physical systems with Lie group symmetries''},
  \emph{Phys.\ Rev.\ A }{\bf 59}, 971 (1999)


     
  \bibitem{Barut:1986}
  A.~Barut and R.~Raczka,
  \emph{``Theory of Group Representations and Applications''},
   World Scientific (1987).


   \bibitem{Altinoluk:2013rua} 
  T.~Altinoluk, C.~Contreras, A.~Kovner, E.~Levin, M.~Lublinsky and A.~Shulkin,
  \emph{``QCD Reggeon Calculus From KLWMIJ/JIMWLK Evolution: Vertices, Reggeization and All''},
  \emph{JHEP} {\bf 1309}, 115 (2013)
  
 

  \bibitem{Kovner:2005en} 
  A.~Kovner and M.~Lublinsky,
  \emph{``From target to projectile and back again: Selfduality of high energy evolution''},
  \emph{Phys.\ Rev.\ Lett.\ } {\bf 94}, 181603 (2005)


  
 \bibitem{Weigert:2000gi} 
  H.~Weigert,
  \emph{``Unitarity at small Bjorken x''},
  \emph{Nucl.\ Phys.\ A } {\bf 703}, 823 (2002)
 
 \bibitem{Blaizot:2002np} 
  J.~P.~Blaizot, E.~Iancu and H.~Weigert,
  \emph{``Nonlinear gluon evolution in path integral form''},
  \emph{Nucl.\ Phys.\ A}  {\bf 713}, 441 (2003)
  
  
  
  \bibitem{nlo:balitsky}  
  I.~Balitsky and G.~A.~Chirilli,
  \emph{``Next-to-leading order evolution of color dipoles''},
  \emph{Phys.\ Rev.\ D}  {\bf 77}, 014019 (2008)

\bibitem{nlo:kovner}
A.~Kovner, M.~Lublinsky and Y.~Mulian,
  \emph{``Jalilian-Marian, Iancu, McLerran, Weigert, Leonidov, Kovner evolution at next to leading order''},
  \emph{Phys.\ Rev.\ D} {\bf 89}, no. 6, 061704 (2014)
  
  
\bibitem{nlo:lublinsky}
M.~Lublinsky and Y.~Mulian,
  \emph{``High Energy QCD at NLO: from light-cone wave function to JIMWLK evolution''},
  \emph{JHEP} {\bf 1705}, 097 (2017)

\bibitem{nlo:caron-huot}
S.~Caron-Huot,
  \emph{``Resummation of non-global logarithms and the BFKL equation''}
  \emph{JHEP} {\bf 1803}, 036 (2018)





 



     
    
 
 
 
  
 
 
 
 
\end{thebibliography}
\end{document}